\documentclass[structabstract]{aa}

\usepackage{amsmath}
\usepackage{amssymb}
\usepackage{txfonts}
\usepackage{booktabs}
\usepackage{array}
\usepackage{graphicx}
\usepackage{dcolumn}
\newcolumntype{L}{D{+}{\,\pm\,}{1,1}}
\usepackage{natbib,twoopt}
\usepackage{color}
\usepackage[breaklinks=true,draft]{hyperref} 
\bibpunct{(}{)}{;}{a}{}{,} 
\newcommandtwoopt{\citeads}[3][][]{\href{http://adsabs.harvard.edu/abs/#3}%
{\citealp[#1][#2]{#3}}}
\newcommandtwoopt{\citepads}[3][][]{\href{http://adsabs.harvard.edu/abs/#3}%
{\citep[#1][#2]{#3}}}
\newcommandtwoopt{\citetads}[3][][]{\href{http://adsabs.harvard.edu/abs/#3}%
{\citet[#1][#2]{#3}}}
\newcommandtwoopt{\citeyearads}[3][][]%
{\href{http://adsabs.harvard.edu/abs/#3}{\citeyear[#1][#2]{#3}}}

\def\ccm{\,{\rm cm^{-3}}}

\def\radm{\,\mathrm{rad\,m^{-2}}}
\def\muG{\,\mu{\rm G}}

\begin{document}

\title{Relics in galaxy clusters at high radio frequencies
	\thanks{Based on observations with the 100-m telescope at Effelsberg,
		      operated by the Max-Planck-Institut f\"ur Radioastronomie (MPIfR) on behalf of the Max-Planck-Gesellschaft.}
}
\author{M.~Kierdorf{\inst{\ref{inst1}}
\thanks{\email{kierdorf@mpifr-bonn.mpg.de}}}
\and R.~Beck\inst{\ref{inst1}}
\and M.~Hoeft\inst{\ref{inst2}}
\and U.~Klein\inst{\ref{inst3}}
\and R.~J.~van Weeren\inst{\ref{inst4}}
\and W.~R.~Forman\inst{\ref{inst4}}
\and C.~Jones\inst{\ref{inst4}}
}

\institute{Max-Planck-Institut f\"ur Radioastronomie, Auf dem H\"ugel 69, 53121 Bonn, Germany\label{inst1}
\and Th\"uringer Landessternwarte, Sternwarte 5, 07778 Tautenburg, Germany\label{inst2}
\and Argelander-Instit\"ut f\"ur Astronomie, Auf dem H\"ugel 71, 53121 Bonn, Germany\label{inst3}
\and Harvard-Smithsonian Center for Astrophysics, 60 Garden Street, Cambridge, MA 02138, USA \label{inst4}
}

\date{Received date / Accepted date}
\titlerunning{Galaxy Cluster Relics at High Radio Frequencies}
\authorrunning{M. Kierdorf et al.}

\abstract
{}
{
	The magnetic properties of radio relics located at the peripheries of galaxy clusters are investigated at high radio frequencies, where the emission is expected to be free of Faraday depolarization. The degree of polarization is a measure of the magnetic field compression and hence the Mach number. Polarization observations can also be used to confirm radio relic candidates.
}
{
	We observed three radio relics in galaxy clusters and one radio relic candidate at 4.85 and 8.35\,GHz in total emission and linearly polarized emission with the Effelsberg 100-m telescope. In addition, we observed one radio relic candidate in X-rays with the Chandra telescope. We derived maps of polarization angle, polarization degree, and Faraday rotation measures.
}
{
	The radio spectra of the integrated emission below 8.35\,GHz can be well fitted by single power laws for all four relics. The flat spectra (spectral indices of 0.9 and 1.0) for the ``Sausage'' relic in cluster CIZA\,J2242$+$53 and the ``Toothbrush'' relic in cluster 1RXS\,06$+$42 indicate that models describing the origin of relics have to include effects beyond the assumptions of diffuse shock acceleration. The spectra of the radio relics in ZwCl\,0008$+$52 and in Abell\,1612 are steep, as expected from weak shocks (Mach number $\approx 2.4$). Polarization observations of radio relics offer a method to measure the strength and geometry of the shock front. We find polarization degrees of more than 50\,\% in the two prominent Mpc-sized radio relics, the Sausage and the Toothbrush, which are among the highest fractions of linear polarization detected in any extragalactic radio source to date. This is remarkable because the large beam size of the Effelsberg single-dish telescope corresponds to linear extensions of about 300\,kpc at 8.35\,GHz at the distances of the relics. The high degree of polarization indicates that the magnetic field vectors are almost perfectly aligned along the relic structure, as expected for shock fronts observed edge-on. The polarization degrees correspond to Mach numbers of $>2.2$. Polarized emission is also detected in the radio relics in ZwCl\,0008$+$52 and, for the first time, in Abell\,1612. The smaller sizes and lower degrees of polarizations of the latter relics indicate a weaker shock and/or an inclination between the relic and the sky plane. Abell\,1612 shows a complex X-ray surface brightness distribution, indicating a recent major merger and supporting the classification of the radio emission as a radio relic.

	In our cluster sample no wavelength-dependent Faraday depolarization is detected between 4.85\,GHz and 8.35\,GHz, except for one component of the Toothbrush relic. Faraday depolarization between 1.38\,GHz and 8.35\,GHz varies with distance from the center of the host cluster 1RXS\,06$+$42, which can be explained by a decrease in electron density and/or in strength of a turbulent (or tangled) magnetic field. Faraday rotation measures show large-scale gradients along the relics, which cannot be explained by variations in the Milky Way foreground.
}
{
	Single-dish telescopes are ideal tools to confirm relic candidates and search for new relic candidates.
	Measurement of the wavelength-dependent depolarization along the Toothbrush relic shows that the electron density of the ICM and strength of the tangled magnetic field decrease with distance from the center of the foreground cluster. Large-scale regular fields appear to be present in intergalactic space around galaxy clusters.
}

\keywords{Galaxies: clusters: general -- galaxies: clusters: intracluster medium -- galaxies: clusters: individual: CIZA J2242+53, 1RXS 06+42, ZwCl 0008+52, Abell 1612 -- magnetic fields -- X-rays: galaxies: clusters}

\maketitle

\section{Introduction}
\label{sec:intro}

	\begin{table*}
	\centering
	\caption{Relic properties}
	\begin{tabular}[ht!]{lcccccccc}
		\toprule\toprule
		Cluster (``Relic'')		                & z 	  & Scale  		         & $L_{\text{X},\,\text{0.1-2.4\,keV}}$& $S_{1372\,\text{MHz}}$ & LLS		&$R_{\rm cc}$  &$\Delta R$& References	\\
		                            &      	& (kpc\,/\,$\arcmin$) & $10^{44}$ erg\,s$^{-1}$				&(mJy)			& (kpc) 		&(kpc)&(kpc)&     		\\
		(1)	   		   & (2)	  & (3)	             	& (4)                   	&(5)			& (6)   		&(7)&(8)& (9)  		\\
		\midrule
		CIZA\,J2242$+$53 (``Sausage'')	&0.192	&	195 & 6.8 & 156		&	2000 &1500 & 60& $A$\\
		1RXS\,06$+$42 (``Toothbrush'') &0.225	&	220 & 7.7 & 320			&	1870 	&1250 & 70& $B$\\
		ZwCl\,0008$+$52 (``ZwCl'')	    &0.103	&	115 &$\sim 0.5$ & 56			&	1400 	&900 & 90& $C$\\
		Abell\,1612 (``A\,1612'')  	    &0.179	&	184 & $1.8$ & 63			&	780				&1300& 130& $D$\\
		\bottomrule&
	\end{tabular}
	\tablefoot{Columns:
		(1) name of the galaxy cluster and radio relic;
		(2) redshift;
		(3) projected linear scale;  
		(4) X-ray luminosity of the cluster;
		(5) flux density of the relic at 1372\,MHz;
		(6) largest linear size (LLS) of the relic;
		(7) distance $R_{\rm cc}$ of relic from cluster center;
		(8) width $\Delta R$ of the relic as estimated from high-resolution radio observations;
		(9) references for values in columns (2), (4), (5), (6), (7), and (8).
		\newline $A$: \citet{2007ApJ...662..224K},  \citet{CIZA}, \cite{2010Sci...330..347V},
		\newline $B$: \citet{2013MNRAS.433..812O}, \cite{toothbrush},
		\newline $C$: \cite{zwicky},
		\newline $D$: this work, \cite{a1612} }
	\label{tab:parameters}
	\end{table*}

	Radio relics are large diffuse radio sources located at the outskirts of galaxy clusters. These sources are not associated with any cluster galaxy, which follows from comparison of radio and optical observations, i.e. the large diffuse radio sources have no optical counterparts. The existence of such radio objects,	observed by synchrotron emission in the radio regime, reveals the presence of relativistic electrons (and maybe a small contribution of positrons) and magnetic fields with field strengths of a few $\mu$G within the intra-cluster medium (ICM). The radio spectra ($I_{\nu}\propto\nu^{-\alpha}$) of cluster relics are steep, with spectral indices $\alpha\,\geq\,1$, see \citet{2012A&ARv..20...54F} for a review.

	The origin of radio relics in galaxy cluster is still a matter of debate. The present idea is that relics trace shock waves which move away from the cluster center towards regions with lower ICM density \citep{ensslin}. The shocks result from mergers of galaxy clusters, predicted in the hierarchical model of structure formation. Radio relics modeled in cosmological cluster merger simulations nicely reproduce many features of observed radio relics \citep{2008MNRAS.391.1511H,2011ApJ...735...96S,2012MNRAS.421.1868V,2015ApJ...812...49H}. Several radio relics are reported to be polarized, some with a very high polarization fraction above 50\,\%  \citep{2009A&A...494..429B,2010Sci...330..347V,zwicky,toothbrush,2012MNRAS.426.1204K,2014ApJ...786...49L,2015MNRAS.453.3483D}. The high degree of polarization could originate from a large-scale regular magnetic field or from compression of a small-scale tangled field in the shock front, which aligns the magnetic field lines along the structure of the shock \citep{1980MNRAS.193..439L}. Relics with a high fractional polarization typically show polarization vectors\footnote{In this paper we are using ``vector'' for the orientation of the magnetic plane of the incoming electromagnetic wave. }
	perpendicular to the relic shape, a fact readily explained via shock compression \citep{ensslin}. In shock waves the particles can be continuously accelerated via diffusive shock acceleration (DSA), which gives rise to a straight and steep radio spectrum (see \citet{2016RPPh...79d6901M} for a recent review). On the other hand, \citet{2016MNRAS.455.2402S} found evidence for spectral breaks in the two most prominent cluster relics, ``Sausage'' and ``Toothbrush''. 
	A break in the spectrum would indicate a more complicated origin of the relativistic electrons, for example re-acceleration of aged seed electrons (e.g., \citet{2005ApJ...627..733M}), or a significant increase of the downstream magnetic field strength within the cooling time of cosmic-ray electrons (CREs) \citep{2016MNRAS.462.2014D}.

	The steep spectra of radio relics hampers the detection of polarized emission at high frequencies. On the other hand, radio relics are known to have high degrees of polarization, which makes them perfect targets to study their polarization and magnetic field properties at high radio frequencies where the effect of Faraday depolarization is small. As high-frequency observations with synthesis radio telescopes suffer from missing short spacings, reducing the visibility of extended emission, single-dish telescopes are preferable.

	We observed three radio relics and one relic candidate with the Effelsberg 100-m single-dish radio telescope at two high radio frequencies, namely at 8.35\,GHz ($\lambda$3.6\,cm) and 4.85\,GHz ($\lambda$6.2\,cm). The slope of the radio spectrum of the integrated emission and the average degree of polarization are used as independent measures of the Mach numbers in the shock fronts of the relics. We measured the Faraday rotation between the two Effelsberg frequencies and we investigated the depolarization between 8.35\,GHz (Effelsberg) and 1.38\,GHz (Westerbork), to obtain information on the distribution of magnetic fields and ionized gas in front of the relics (i.e. the ICM and/or the Milky Way). General information on the observed relics is summarized in Table \ref{tab:parameters}.
	


	The layout of this paper is as follows. An overview of the observations and data reduction is given in Section~\ref{sec:obs}. The results obtained from the spectral index, polarization, Faraday depolarization, and Faraday rotation data, and estimates of the magnetic field strengths in the relics are described in Section~\ref{sec:results}, followed by conclusions and discussion in Section~\ref{sec:conclusion}. Throughout this paper, we assume a cosmology with $H_0=69$\,km$^{-1}$\,s$^{-1}$\,Mpc, $\Omega_{\text{M}}=0.3$ and $\Omega_{\lambda}=0.7$.


\section{Observations and data reduction}
\label{sec:obs}

	\begin{table}
	\centering
	\caption{8.35\,GHz observations}
	  \begin{tabular}[t!]{lccccc}
	    \toprule\toprule
	Relic 	& Beam			& Area		& N 		& $\sigma_{\text{I}}$	& $\sigma_{\text{UQ}}$		\\
	name	    & size	        & 			& I; U/Q	& 	& 	\\
	(1)      	&(2)		&(3)        &(4)		&(5)			    &(6)			\\
	  \midrule
	Sausage		&  $ 90\arcsec$	&  $10\arcmin\times 14\arcmin$	&46;\,46	&400		&70\\
	Toothbrush	&  $ 90\arcsec$	&  $10\arcmin\times 15\arcmin$	&40;\,20	&500		&130\\
	ZwCl    	&  $ 90\arcsec$	&  $15\arcmin\times 15\arcmin$	&63;\,36	&200		&60\\
	A\,1612		&  $ 90\arcsec$	&  $15\arcmin\times 15\arcmin$	&34;\,17	&500		&100\\
	    \bottomrule
	  \end{tabular}
	   \tablefoot{Observation parameters:\newline(1) Source name; (2) beam size; (3) map extension (RA $\times$ DEC); (4) number of maps in Stokes I, U, and Q; (5) rms noise in Stokes I in $\mu$Jy/beam; (6) average rms noise in Stokes U and Q in $\mu$Jy/beam.}
	   \label{tab:observation3cm}
	\end{table}
	
	\begin{table}
	  \centering
	\caption{4.85\,GHz observations}
	  \begin{tabular}[h]{lccccc}
	    \toprule\toprule
	 Relic	& Beam 	& Area	& N		& $\sigma_{\text{I}}$	& $\sigma_{\text{UQ}}$	\\
	name	    & size	& 		& I; U/Q	& 	& 		\\
	(1)     &(2)	&(3)  	&(4)		&(5)				&(6)				\\
	  \midrule
	Sausage		& $159\arcsec$	& $16\arcmin\times 26\arcmin$	&32;\,32	&800	&120	\\
	Toothbrush	& $159\arcsec$	& $20\arcmin\times 30\arcmin$	&12;\,12	&1000	&90	\\
	ZwCl		& $159\arcsec$	& $25\arcmin\times 35\arcmin$	&38;\,38	&750	&70	\\
	A\,1612		& $159\arcsec$	& $15\arcmin\times 25\arcmin$	&12;\,12	&500	&80	\\
	    \bottomrule
	  \end{tabular}
	   \tablefoot{Observation parameters:\newline(1) Source name; (2) beam size; (3) map extension (AZM $\times$ ELV); (4) number of maps in Stokes I, U, and Q;  (5) rms noise in Stokes I in $\mu$Jy/beam; (6) average rms noise in Stokes U and Q in $\mu$Jy/beam.}
	   \label{tab:observation6cm}
	 \end{table}

	The radio continuum observations were performed with the Effelsberg 100-m telescope in October/November 2010, January 2011, October 2011 and August 2014 using the $\lambda$3.6\,cm (8.35\,GHz with 1.1\,GHz bandwidth) single-horn and $\lambda$6.2\,cm (4.85\,GHz with 0.5\,GHz bandwidth) dual-horn receiving systems, providing data channels in Stokes I, U and Q for each horn. We used a digital broadband polarimeter as backend. The radio sources 3C\,48, 3C\,138, 3C\,147, 3C\,286  and 3C\,295 were observed as flux density and pointing calibrators, using the flux densities from \citet{2000A&AS..145....1P}. We adopted polarization angles of $-\,11\degr$ and $+\,33\degr$ for 3C\,138 and 3C\,286, respectively \citep{2013ApJS..206...16P}. At 8.35\,GHz the maps were scanned alternating in RA and DEC directions, at 4.85\,GHz in azimuthal direction only. The main part of data reduction, combining the scans into a map and subtracting the baselevels, was accomplished using the \textit{NOD2}-based software package \citep{nod2}. Thereby each map in I, U and Q was corrected for severe scanning effects, Radio Frequency Interference (RFI) and incorrect baselevels. The dual-horn maps at 4.85\,GHz were combined using software beam-switching \citep{1979A&A....76...92E} and transformed into the RA, DEC coordinate system. The level of instrumental polarization in Effelsberg observations is below 1\% and was not corrected for. All maps in Stokes parameters I, U and Q were averaged and combined using the basket-weaving method \citep{emerson}. The final maps were smoothed by about 10\% from originally 82\arcsec\ to 90\arcsec\ (1.50\arcmin) at 8.35\,GHz and from originally 146\arcsec\ to 159\arcsec\ (2.65\arcmin) at 4.85\,GHz.

	A summary of the radio observations is given in Tables~\ref{tab:observation3cm} and \ref{tab:observation6cm}. The rms noise in the maps of U and Q are similar, so that only the average value is given. The rms noise in the U and Q maps is several times lower than that in Stokes I because (a) the total intensity is much more affected by scanning effects due to clouds and (b) the confusion level of weak sources is reached in most Stokes I maps, while the confusion level in U and Q is at least one order of magnitude lower and is not reached by our observations.

	The Astronomical Image Processing System (\textit{AIPS}) was used to compute the radio continuum maps with respect to polarization and magnetic fields properties (i.e. polarized intensity, degree of polarization, polarization angle, and Faraday rotation). The polarized intensity was corrected for positive bias due to noise \citep{1974ApJ...194..249W} with help of the option \textit{POLC} in the task \textit{COMB}. As the noise distribution in the maps of polarized intensity is non-Gaussian, no rms noise values can be given, and the same rms noise levels as those in Stokes U and Q are adopted.


	We also observe Abell\,1612 in X-rays with Chandra for 29\,ks (ObsID 15190) with ACIS-I. The data were reduced with the CIAO task {\tt chav}, following the process explained in \cite{2005ApJ...628..655V}. We used the CALDB~4.7 calibration files. Periods with high background rates and readout artifacts were removed from the data. The data were corrected for the time dependence of the charge transfer inefficiency and gain. We then subtracted the background, using the standard blank sky background files, and made a exposure corrected image in the 0.5--2.0\,keV band, using a pixel binning of 4 (i.e., 2\arcsec~pixel$^{-1}$).

\section{Results}
\label{sec:results}

	\begin{figure*}[htbp]
	\centering
	   \includegraphics[width=9cm, trim=1cm 7cm 0.5cm 7cm, clip=true]{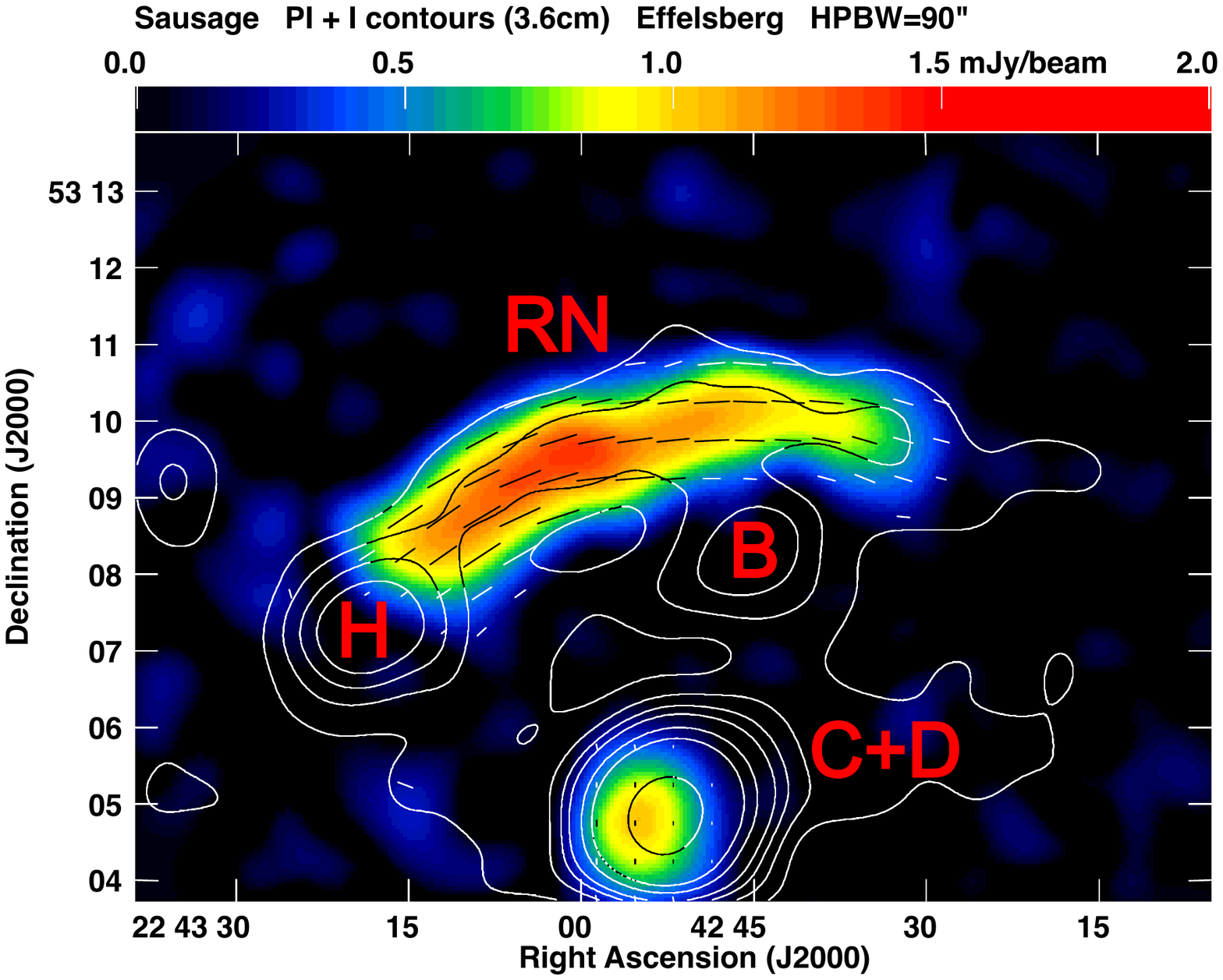}
	   \hspace*{0mm}
	   \includegraphics[width=9cm, trim=1cm 7cm 0.5cm 7cm, clip=true]{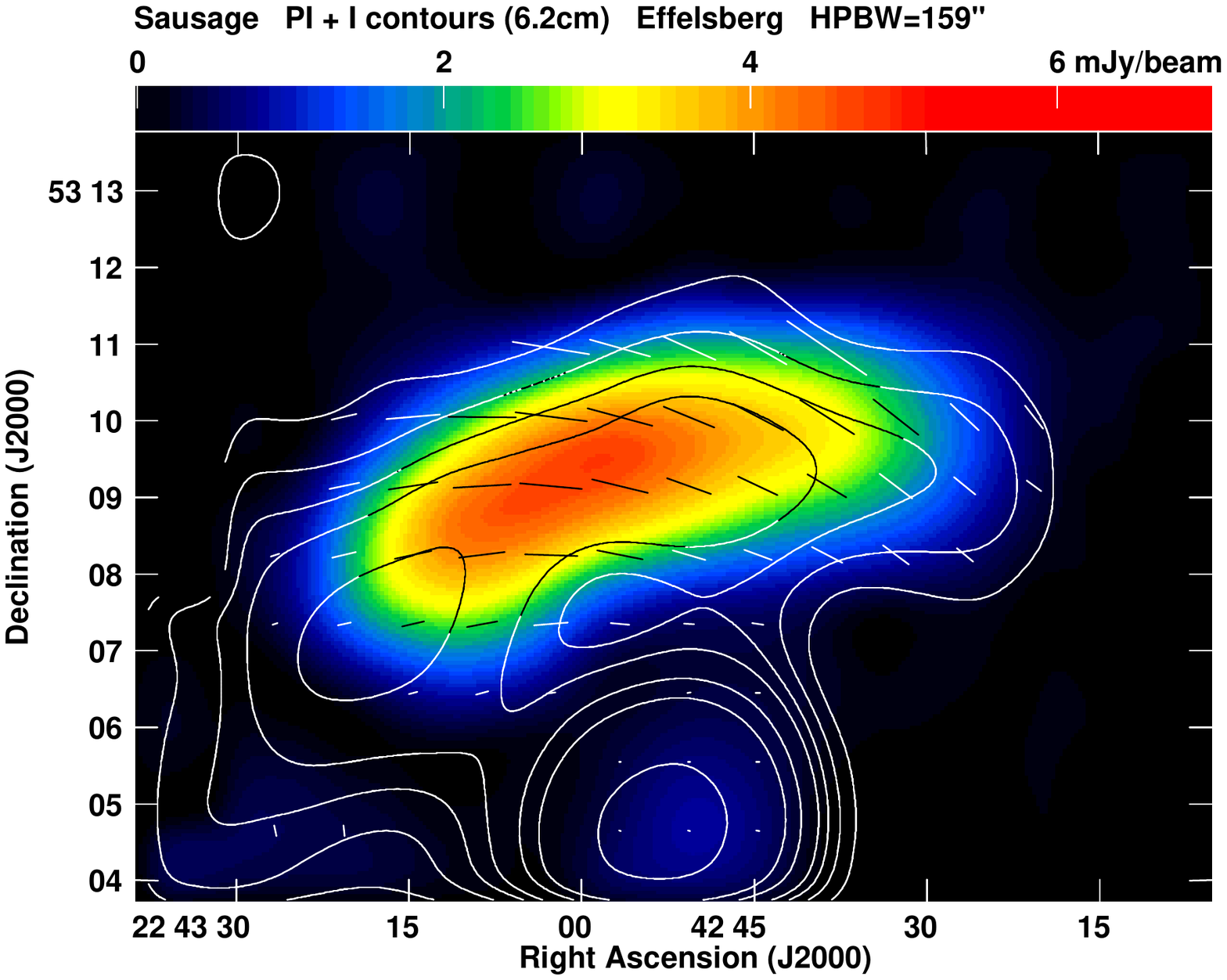}
	     \caption{The polarized emission (color) of the Sausage relic in CIZA\,J2242$+$53, overlaid with total intensity
	     contours.
	     Vectors depict the polarization B--vectors, not corrected for Faraday rotation, with lengths representing
	     the degree of polarization.
	     \textit{Left:} 8.35\,GHz map. Total intensity contours are drawn at levels of [1, 2, 3, 4, 6, 8, 16]\,$\times\,1.2$\,mJy\,/\,beam.
	     A vector length of $1\arcmin$ corresponds to 90\,\% fractional polarization. The beam size is $90\arcsec\times90\arcsec$. The sources are labeled according to \cite{CIZA}.
	     \textit{Right:} 4.85\,GHz map. Total intensity contours are drawn at levels of [1, 2, 3, 4, 6, 8, 16]\,$\times\,2.4$\,mJy\,/\,beam. A vector length of $1\arcmin$ corresponds to 50\,\% fractional polarization.
	     The beam size is $159\arcsec\times159\arcsec$.}
	     \label{fig:sausage}
	\end{figure*}

	\begin{figure*}[htbp]
	\centering
	   \includegraphics[width=10cm, trim=1cm 6cm 0.5cm 7cm, clip=true]{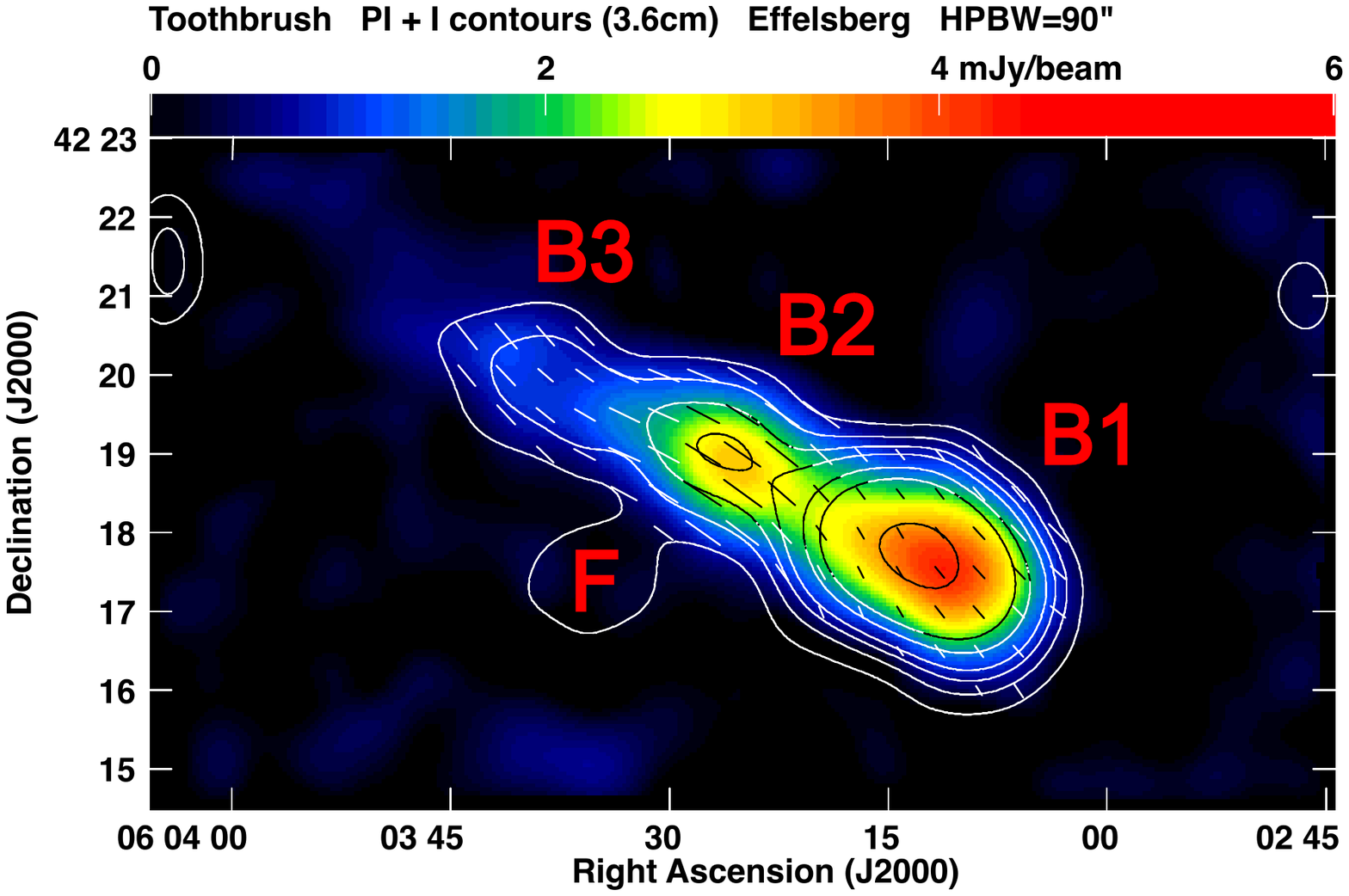}
	   \hspace*{0mm}
	   \includegraphics[width=8cm, trim=1cm 5cm 0.5cm 4cm, clip=true]{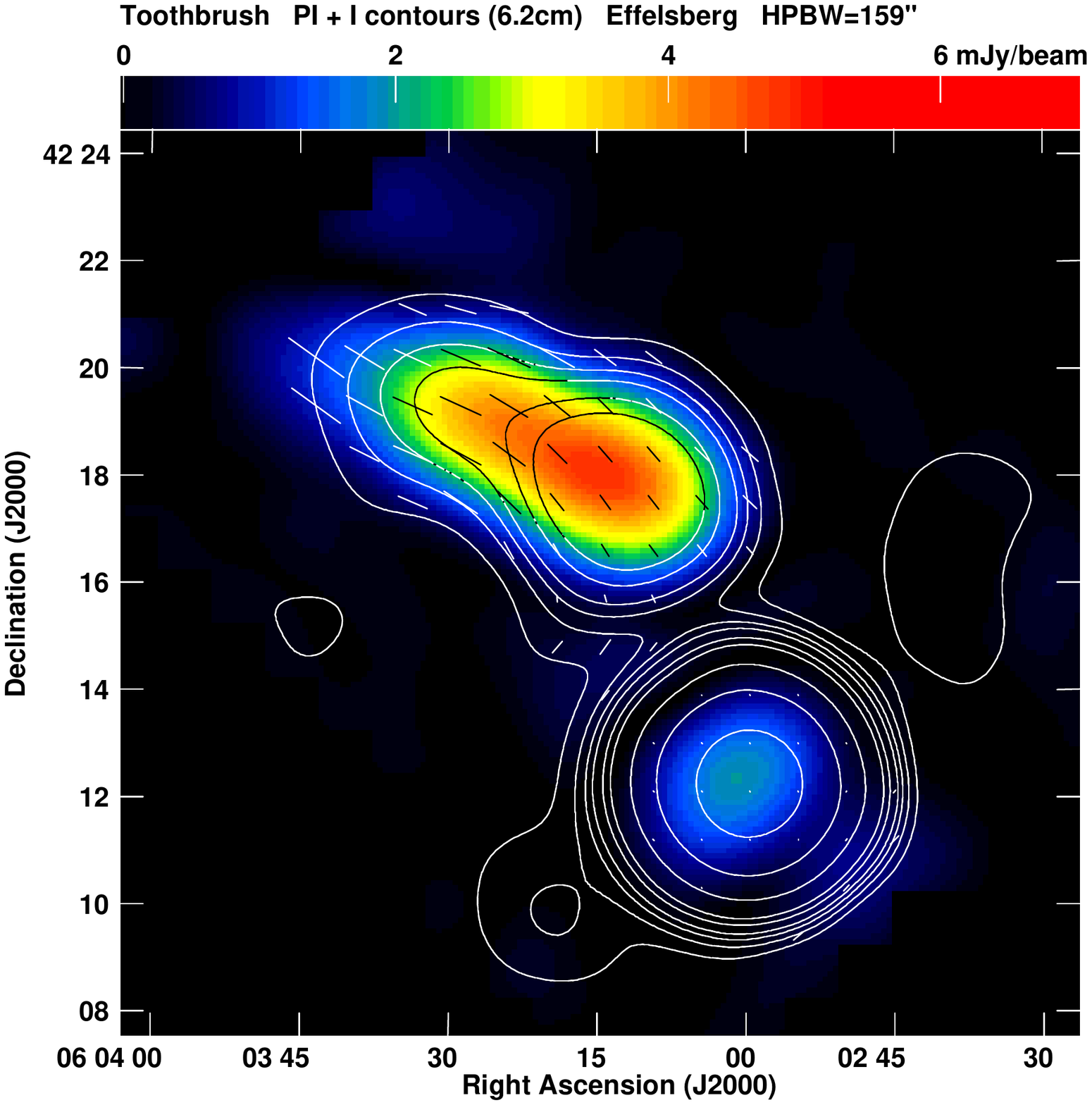}
	     \caption{The polarized emission (color) of the Toothbrush relic in 1RXS\,06$+$42, overlaid with total intensity
	     contours.
	     Vectors depict the polarization B--vectors, not corrected for Faraday rotation, with lengths representing
	     the degree of polarization.
	     \textit{Left:} 8.35\,GHz map. Total intensity contours are drawn at levels of [1, 2, 3, 4, 6, 8, 16]\,$\times\,1.5$\,mJy\,/\,beam.
	     A vector length of $1\arcmin$ corresponds to 90\,\% fractional polarization.
	     The beam size is $90\arcsec\times90\arcsec$.  The sources are labeled according to \cite{toothbrush}.
	     \textit{Right:} 4.85\,GHz map. Total intensity contours are drawn at levels of [1, 2, 3, 4, 6, 8, 16]\,$\times\,3$\,mJy\,/\,beam.
	     A vector length of $1\arcmin$ corresponds to 30\,\% fractional polarization.
	     The beam size is $159\arcsec\times159\arcsec$.}
	     \label{fig:tooth}
	\end{figure*}

	\begin{figure*}[htbp]
	\centering
	   \includegraphics[width=9cm, trim=1cm 7cm 1cm 5cm, clip=true]{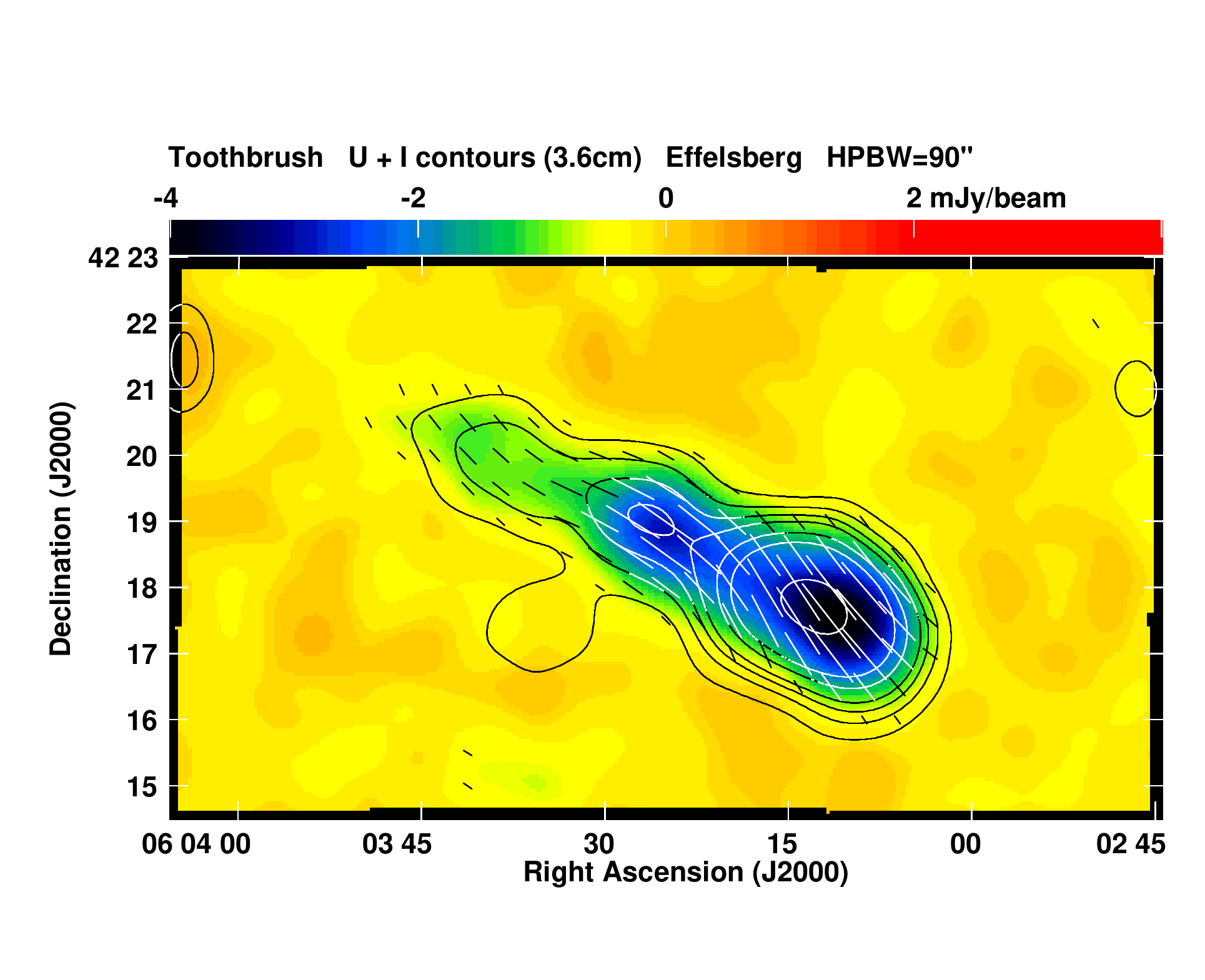}
	   \hspace*{0mm}
	   \includegraphics[width=9cm, trim=1cm 7cm 1cm 5cm, clip=true]{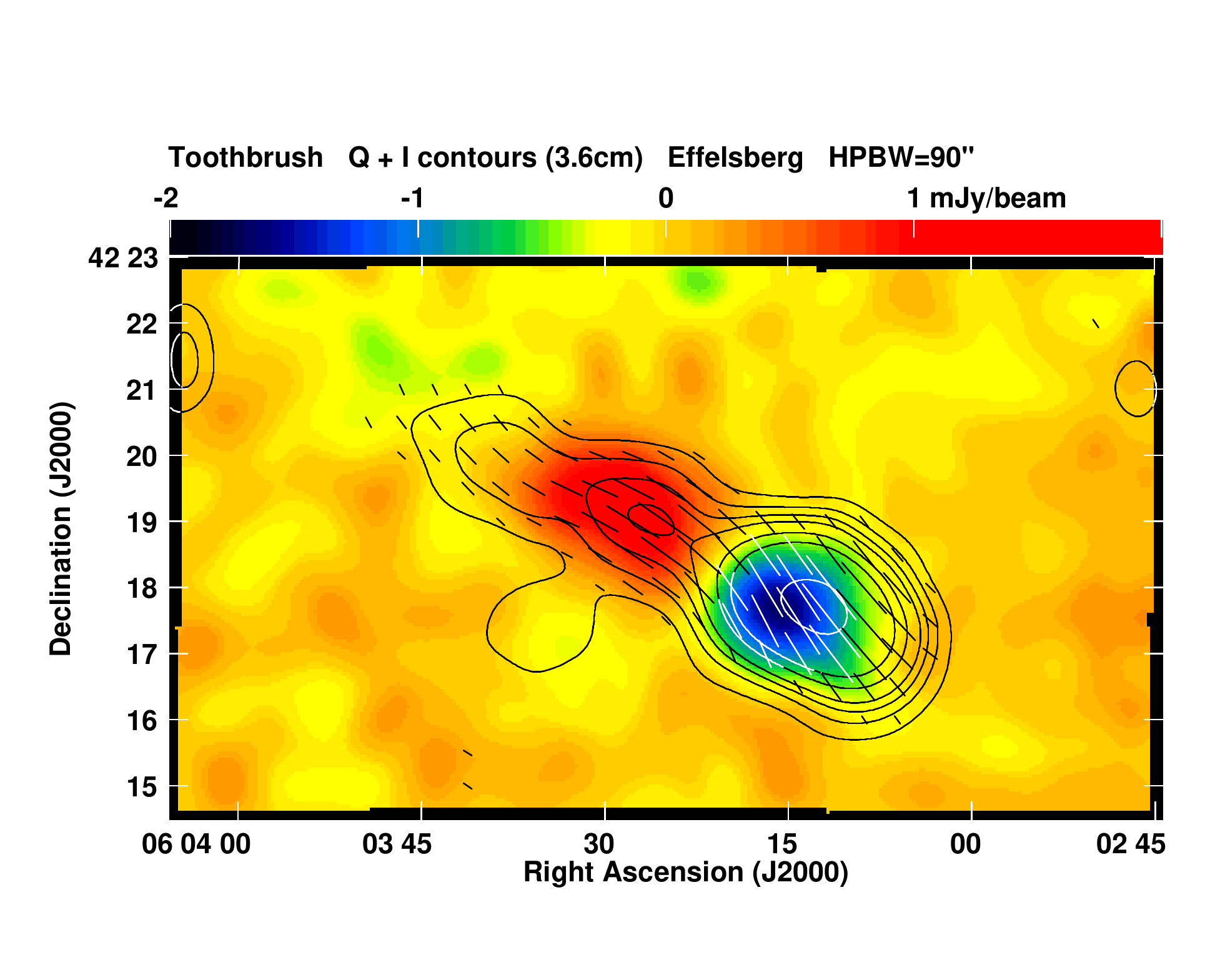}
	     \caption{The polarized emission (color) of the Toothbrush relic 1RXS\,06$+$42 at 8.35\,GHz in Stokes U
	     (\textit{left}) and Stokes Q (\textit{right}), overlaid with total intensity contours.
	     Vectors depict the polarization B--vectors, not corrected for Faraday rotation, with lengths representing the polarized intensity.
	     ($1\arcmin \equiv 3$\,mJy\,/\,beam). Total intensity contours are drawn at levels of
	     [1, 2, 3, 4, 6, 8, 16]\,$\times\,1.5$\,mJy\,/\,beam. The beam size is $90\arcsec\times90\arcsec$. }
	     \label{fig:qu}
	\end{figure*}

	\begin{figure*}[htbp]
	\centering
	   \includegraphics[width=9cm, trim=1cm 6cm 1cm 5cm, clip=true]{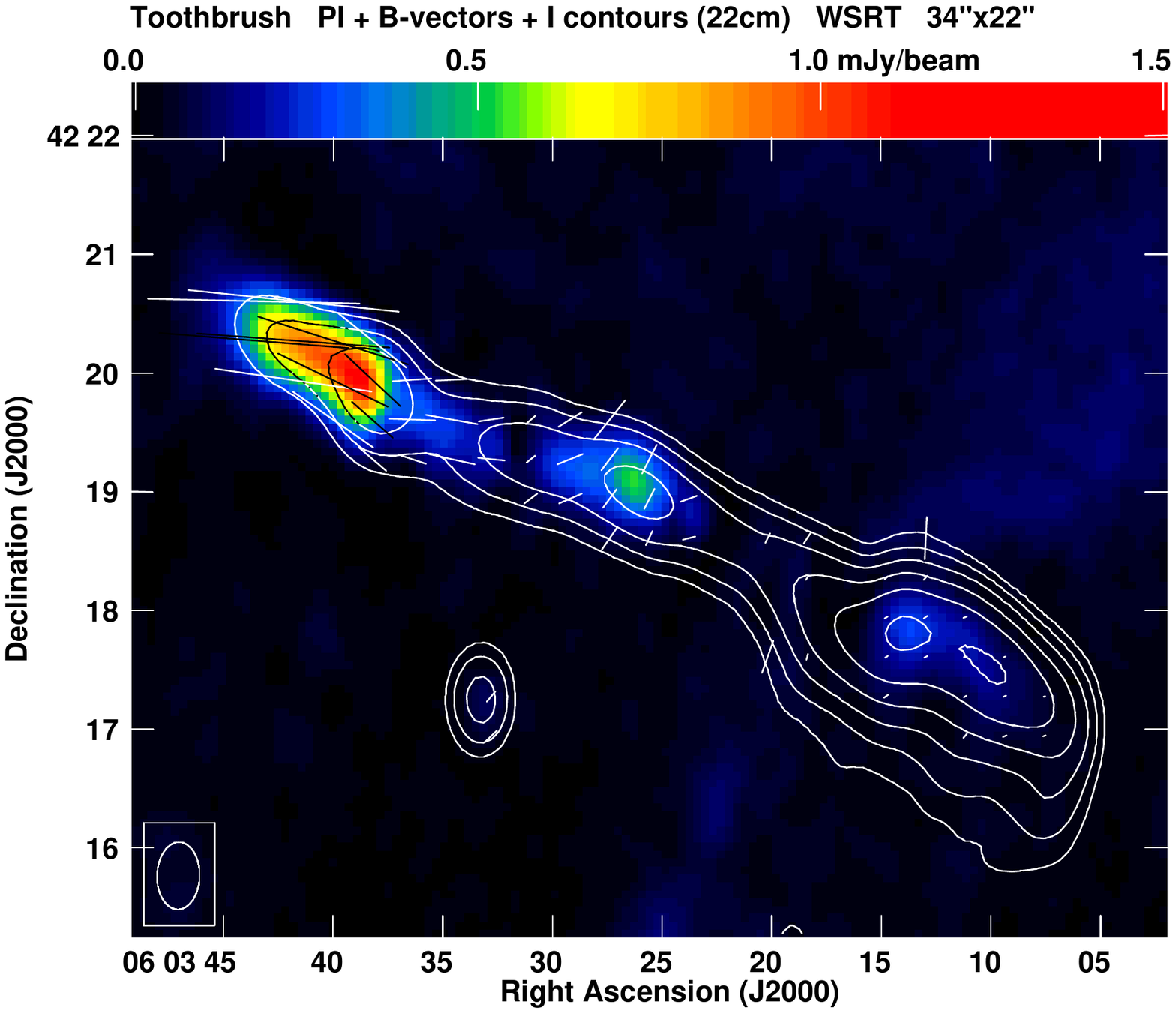}
	   \hspace*{0mm}
	   \includegraphics[width=9cm, trim=1cm 6cm 0.5cm 5cm, clip=true]{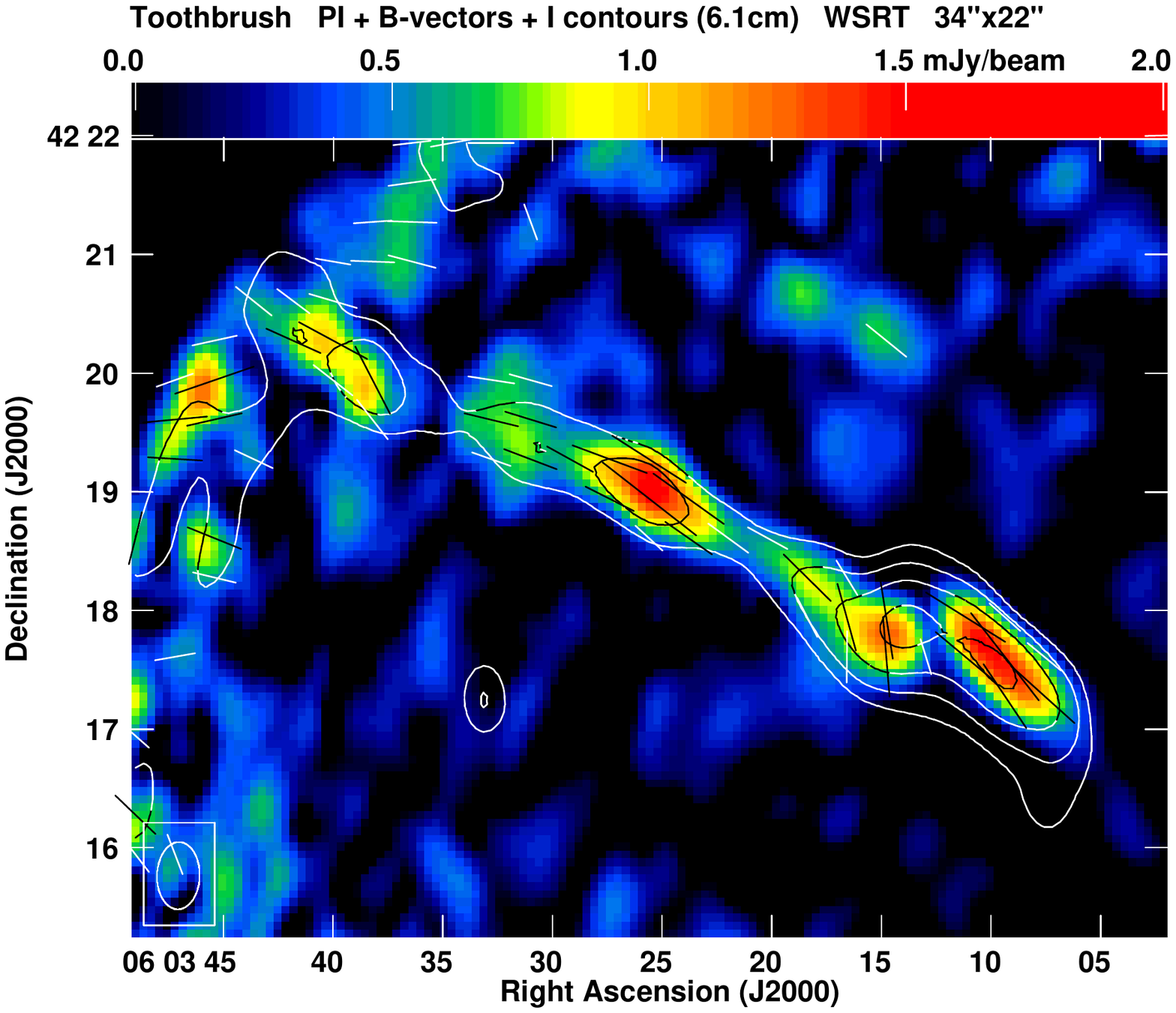}
	     \caption{The polarized emission (color) of the Toothbrush relic in 1RXS\,06$+$42, overlaid with total intensity
	     contours, from observations at the Westerbork Radio Synthesis Telescope (WSRT) \citep{toothbrush}.
	     Vectors depict the polarization B--vectors, not corrected for Faraday rotation, with lengths representing the degree of polarization.
	     The beam size is $22\arcsec\times34\arcsec$.
	     \textit{Left:} 1.38\,GHz map. Total intensity contours are drawn at levels of [1, 2, 4, 6, 8, 16, 32]\,$\times\,1$\,mJy\,/\,beam. A vector length of $1\arcmin$ corresponds to 20\,\% fractional polarization.
	     \textit{Right:} 4.9\,GHz map. Total intensity contours are drawn at levels of [1, 2, 4, 6, 8, 16, 32]\,$\times\,1$\,mJy\,/\,beam. A vector length of $1\arcmin$ corresponds to 75\,\% fractional polarization.
The rms noise increases with increasing distance from the center of the primary beam which is obvious at the eastern edge of the map.
}
	     \label{fig:tooth_WSRT}
	\end{figure*}

	\begin{figure*}[htbp]
	\centering
	   \includegraphics[width=9cm, trim=1cm 4cm 1cm 4cm, clip=true]{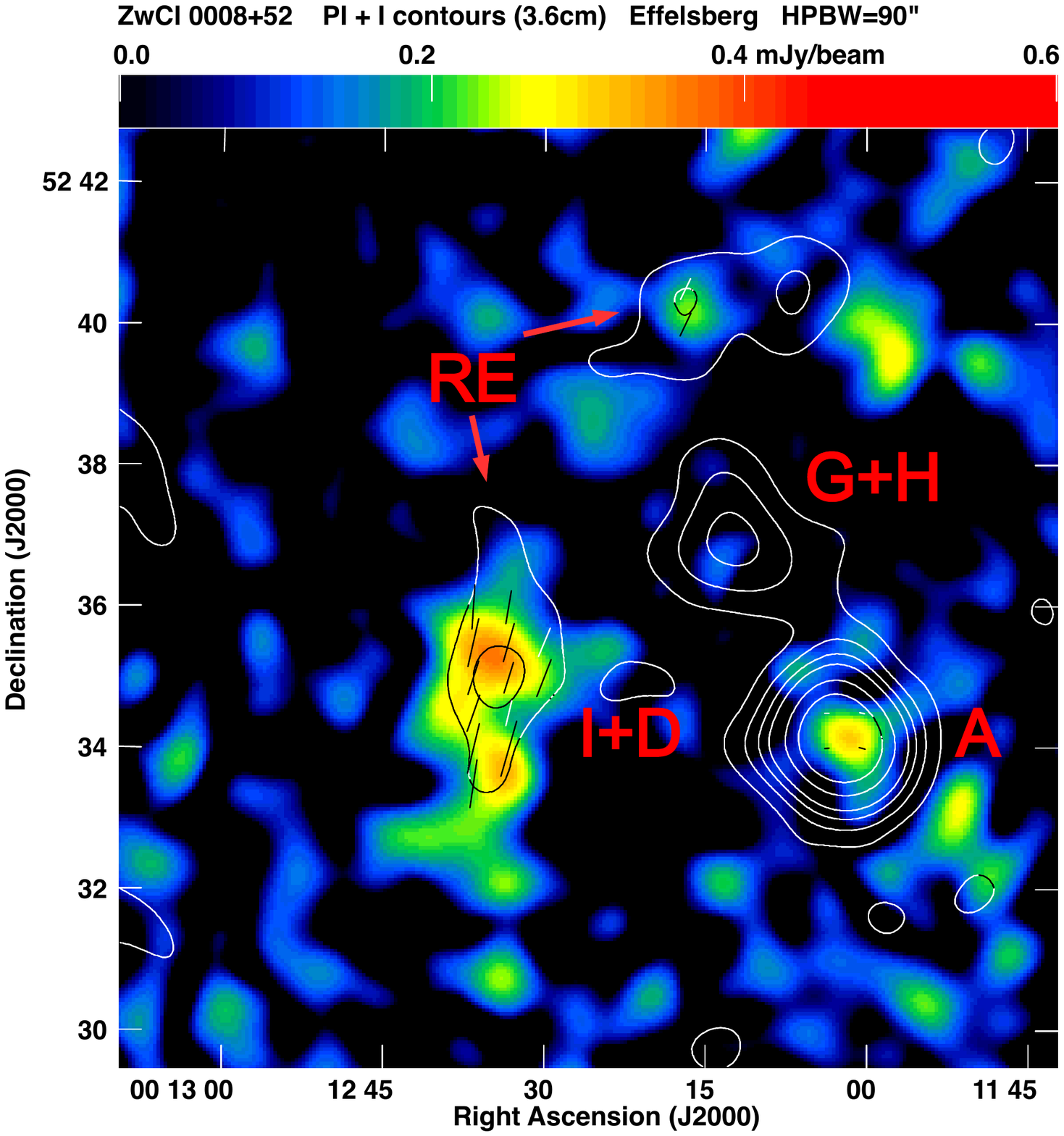}
	   \hspace*{0mm}
	   \includegraphics[width=9cm, trim=1cm 4cm 1cm 4cm, clip=true]{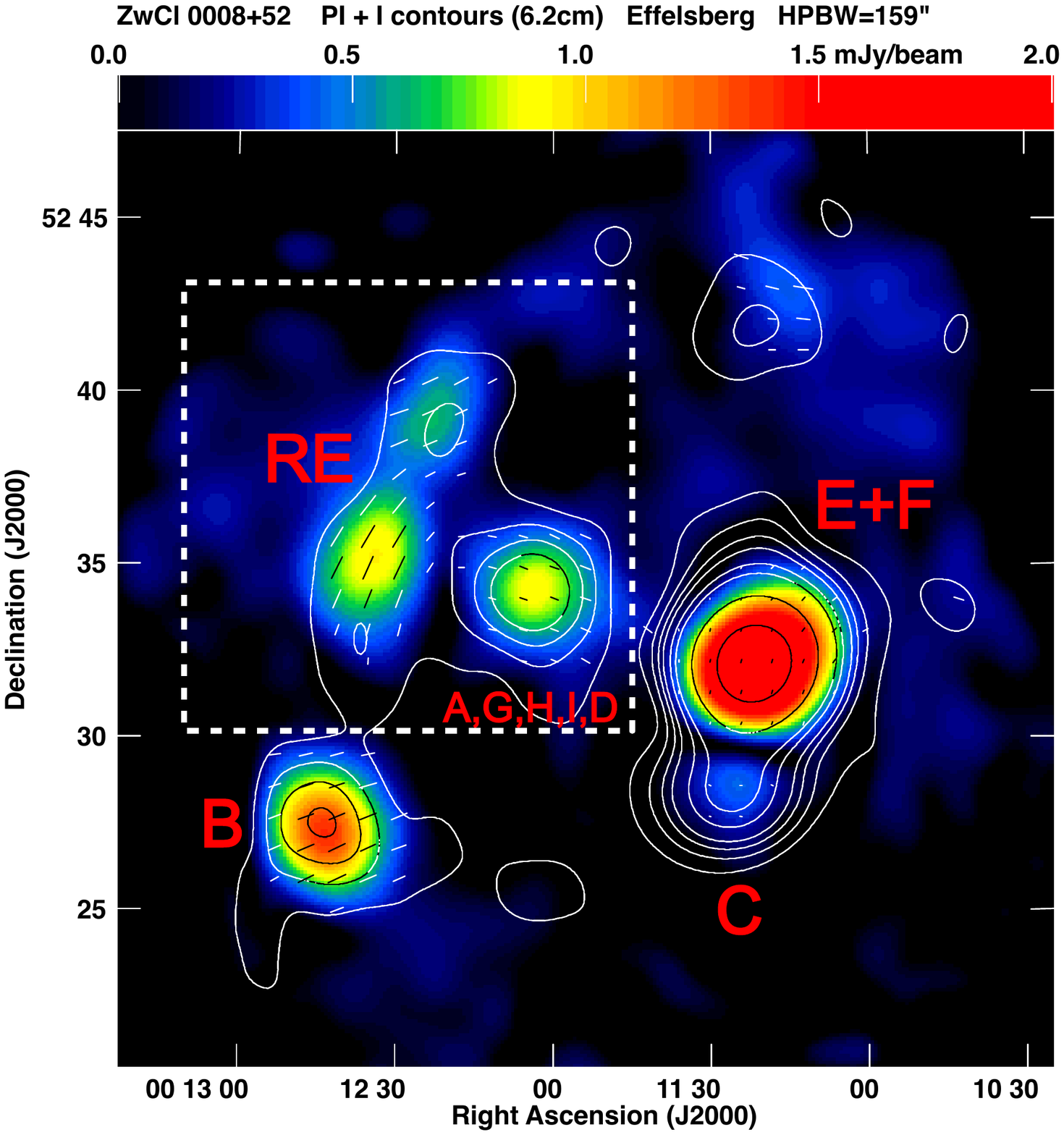}
	     \caption{The polarized emission (color) of the radio relic in ZwCl\,0008$+$52, overlaid with total intensity
contours. Vectors depict the polarization B--vectors, not corrected
	     for Faraday rotation, with lengths representing the degree of polarization.  The sources are labeled according to \cite{zwicky}.
	     \textit{Left:} 8.35\,GHz map. Total intensity contours are drawn at levels of [1, 2, 3, 4, 6, 8, 16]\,$\times\,0.6$\,mJy\,/\,beam. A vector length of $1\arcmin$ corresponds to 50\,\% fractional polarization.
	     The beam size is $90\arcsec\times90\arcsec$.
	     \textit{Right:} 4.85\,GHz map. Total intensity contours are drawn at levels of [1, 2, 3, 4, 6, 8, 16]\,$\times\,2.25$\,mJy\,/\,beam. A vector length of $1\arcmin$ corresponds to 30\,\% fractional polarization.
	     The beam size is $159\arcsec\times159\arcsec$. --
	     Note that the map area of the 8.35\,GHz measurement (left panel) only includes a small region around the eastern relic (white box in the right panel),
	     whereas the map of the 4.85\,GHz measurement (right panel) encompasses the entire cluster.}
	     \label{fig:zwcl}
	\end{figure*}

\begin{figure*}[htbp]
	\centering
	   \includegraphics[width=9.5cm, trim=1cm 5cm 1cm 4cm, clip=true]{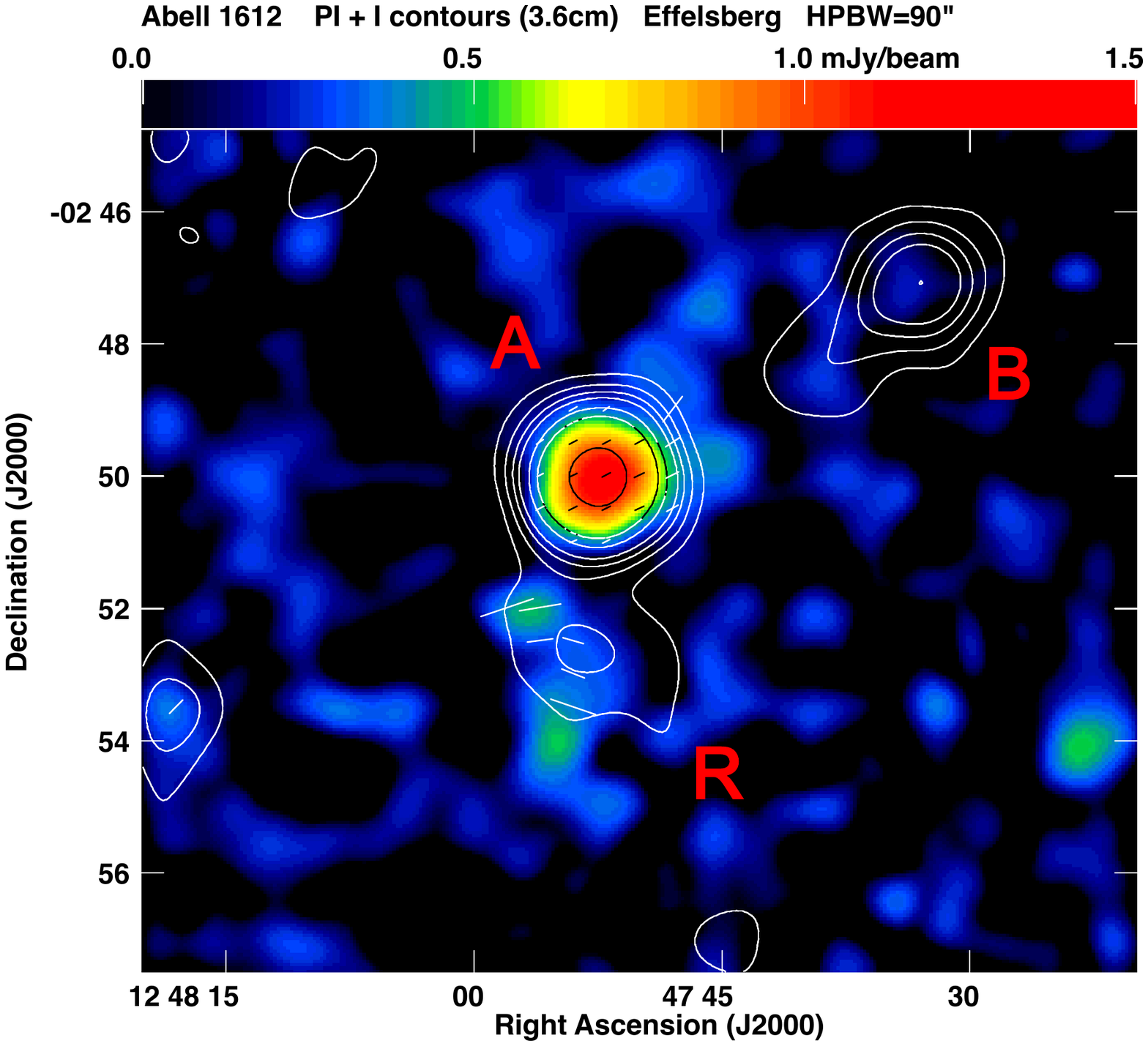}
	   \hspace*{0mm}
	   \includegraphics[width=8.5cm, trim=1cm 3.5cm 1cm 4cm, clip=true]{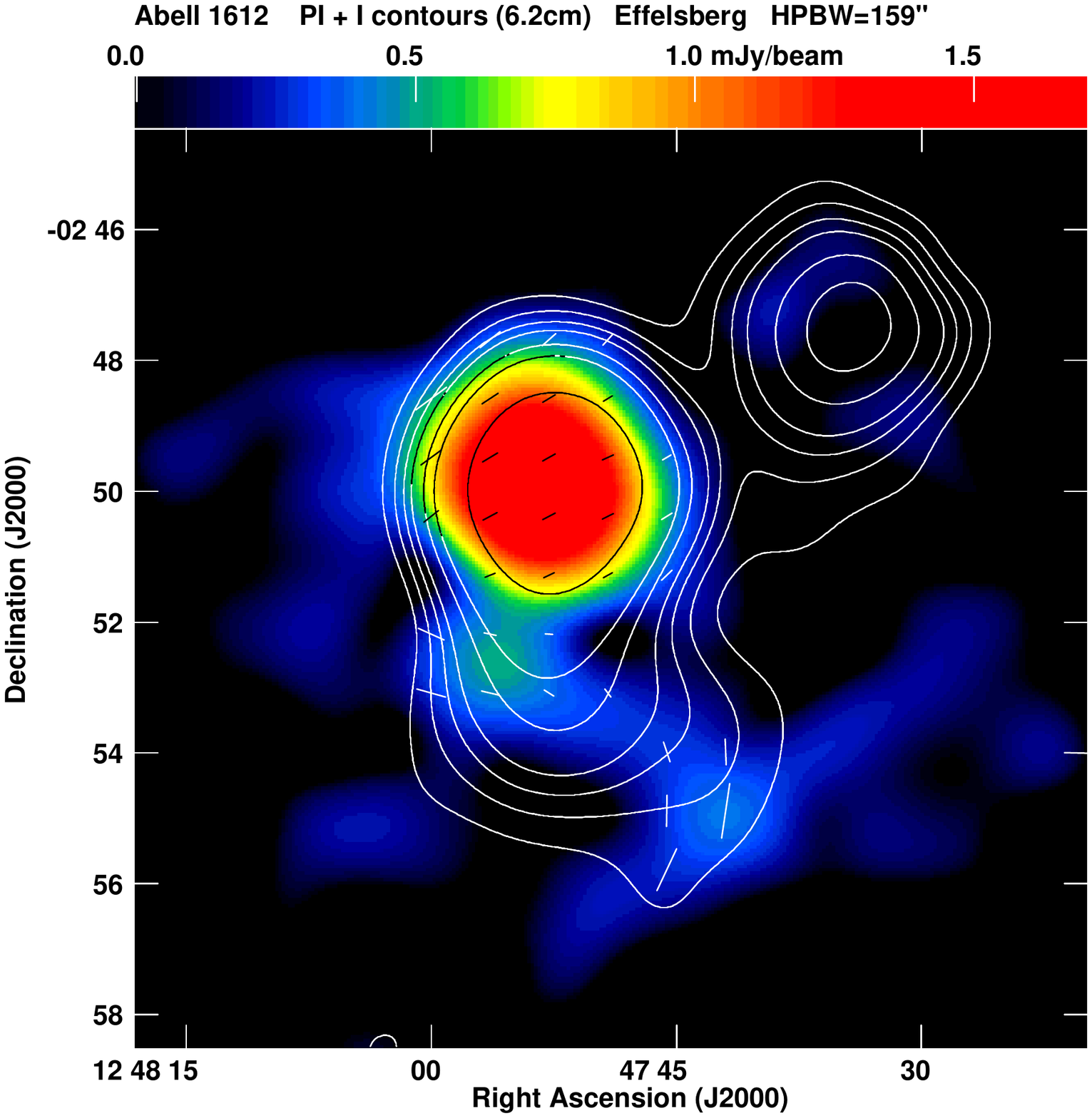}
	     \caption{The polarized emission (color) of the radio relic in Abell\,1612, overlaid with total intensity contours.
	     Vectors depict the polarization B--vectors, not corrected for Faraday rotation, with lengths representing the degree of polarization.
	     \textit{Left:} 8.35\,GHz map. Total intensity contours are drawn at levels of [1, 2, 3, 4, 6, 8, 16]\,$\times\,1.5$\,mJy\,/\,beam. A vector length of $1\arcmin$ corresponds to 30\,\% fractional polarization.
	     The beam size is $90\arcsec\times90\arcsec$.
	     \textit{Right:} 4.85\,GHz map. Total intensity contours are drawn at levels of [1, 2, 3, 4, 6, 8, 16]\,$\times\,1.5$\,mJy\,/\,beam. A vector length of $1\arcmin$ corresponds to 20\,\% fractional polarization.
	     The beam size is $159\arcsec\times159\arcsec$.}
	     \label{fig:a1612}
	\end{figure*}

	The radio continuum images of the four clusters are presented in Figures~\ref{fig:sausage}--\ref{fig:a1612}. In all figures, the lowest contour of total intensity is plotted at the level of 3 times the rms noise in Stokes I. Significant polarized emission is detected in all clusters. The polarization vectors (rotated by $90\degr$) are plotted only if the intensities are above 3 times the average rms noise in Stokes I, U, and Q.

	The orientations of the polarization vectors are rotated by $90\degr$ to show the approximate orientations of the magnetic field component in the plane of the sky. The length of the vectors corresponds to the fractional polarization. Faraday rotation is not corrected in Figures~\ref{fig:sausage}--\ref{fig:a1612}, to show the amount of rotation between the two frequencies. For Faraday rotation measures (RMs) of $|\text{RM}|\,<\,200\radm$ (see Figure~\ref{fig:RM}), the absolute value of the Faraday rotation angle is smaller than $\pm\,45\degr$ at 4.85\,GHz and smaller than $\pm\,15\degr$ at 8.35\,GHz.


\subsection{CIZA\,J2242$+$53 (``Sausage'')}

	In the galaxy cluster CIZA J2242$+$53, \cite{2010Sci...330..347V} detected a 2\,Mpc-sized radio relic which is located in the northern outskirts of the cluster. The X-ray surface brightness distribution clearly shows that the cluster is undergoing a merger \citep{1999A&A...349..389V,2013MNRAS.429.2617O,2014MNRAS.440.3416O}. The 8.35\,GHz and 4.85\,GHz images of the relic obtained in this work are shown in Figure~\ref{fig:sausage}. The sources are labeled according to \cite{CIZA}. We show here the large radio relic (RN) in the northern outskirts of the galaxy cluster. The bright point-like source seen in the bottom part of the images (C+D) is a bright head-tail radio galaxy which is hosted by an elliptical galaxy. The source located east of the relic (H) has a much higher surface brightness than the relic and is also a tailed radio galaxy \citep{CIZA}. The diffuse emission seen in total intensity south of the relic is probably an artifact of the inhomogeneous baselevel (see Section~\ref{sec:flux}).

	Comparison of total (contours) and polarized intensity (color scale) of the image at 8.35\,GHz with better resolution (Figure~\ref{fig:sausage} left) implies that the point source at the eastern end of the relic (H) is not polarized and hence not part of the relic.

	The magnetic field vectors show a well defined pattern aligned along the relic (RN). This indicates that the magnetic field component in the plane of the sky within the relic is well ordered, presumably because turbulent magnetic fields became anisotropic by the merger shock (see Section \ref{sec:mach}). The degree of polarization at 8.35\,GHz decreases from about $\approx$\,55\,\% at the eastern end of the relic to about $\approx$\,25\,\% at the western end (Figure~\ref{fig:percpol} left), similar to the polarization fraction measured at 1.4\,GHz by \citet{2010Sci...330..347V} with a much higher angular resolution of $5.2\arcsec \times 5.1\arcsec$. This suggests that (1) the magnetic field is ordered (anisotropic) on scales larger than the Effelsberg beam (corresponding to about 300\,kpc at 8.35\,GHz for this relic), and (2) wavelength-dependent Faraday depolarization in the cluster medium in front of the relic plays little role at these frequencies, in contrast to the Toothbrush relic (see Chapter \ref{sec:dp}).

	The average degree of linear polarization in the eastern part is $53 \pm 13 \,\%$. The value is taken from the mean of all pixels in an area of about one beam and the error from the rms noise of the images in polarized and total intensity. {\em This is among the highest values measured in radio sources}, comparable to the highly polarized shock fronts of old supernova remnants \citep[e.g.][]{2007A&A...470..969X}. \citet{1998A&A...331..475F}, using the Effelsberg telescope, showed that linear polarization levels of 50\% are common in tailed radio galaxies. Similar levels have been also observed in the lobes of giant radio galaxies, e.g. NGC\,6251 \citep{1996IAUS..175..317M}. More recently, \citet{2016MNRAS.461.3516M}, using the Sardinia Radio Telescope, measured a polarization level as high as 70\% at 6.6\,GHz in the head-tail radio galaxy 3C\,129.

	The degrees of polarization measured from our Effelsberg maps at 8.35\,GHz and 4.85\,GHz at 159\arcsec\ resolution are similar. The maximum degree of polarization is about 45\,\%, smaller than at $90\arcsec$ resolution because the curvature of the relic's magnetic field within the large beam may lead to depolarization.

\subsection{1RXS\,06$+$42 (``Toothbrush'')}

	Some characteristics of the famous Toothbrush radio relic in the galaxy cluster 1RXS06$+$42 are summarized in \citet{toothbrush,2016ApJ...818..204V}: The large bright radio source has a LLS of 1.87\,Mpc (Table \ref{tab:parameters}). The source is known to be polarized and the spectral index gradient expected for radio relics has been clearly observed for the Toothbrush, justifying the classification of the source as a relic. Enigmatically, the relic does not follow the X-ray surface brightness but extends into a region with very low ICM density towards the east. A similar behavior was also observed in the spectacular giant radio relic hosted in the double X-ray peak galaxy cluster Abel\,115 (i.e. \citet{2016MNRAS.460L..84B}).

	Figure~\ref{fig:tooth} shows the 8.35\,GHz and 4.85\,GHz images of the Toothbrush radio relic obtained in this work. The 8.35\,GHz image shows only the upper part of the galaxy cluster with the bright extended radio relic. The 4.85\,GHz image shows a larger region around the relic including a strong background source in the south-west part of the cluster. According to observations with higher resolution by \citet{toothbrush}, shown in Figure~\ref{fig:tooth_WSRT}, the relic consists of a bright western part and a fainter linear extension to the northeast, which is also seen in the Effelsberg images. The relic apparently comprises three components, following \citet{toothbrush}, we label the components B1, B2, and B3, see Figure~\ref{fig:tooth}.

	Our measurements show that the relic is polarized over its entire length. The polarization fraction at 8.35\,GHz at $90\arcsec$ resolution, see Figure~\ref{fig:percpol} right panel, strongly varies along the relic, with $30\pm8\%$ in region B3, a maximum of $45\pm7\%$ in region B2, and $15\pm1\%$ in region B1. Our results of the degrees of polarization for all three components are similar to those obtained with the WSRT at 4.9\,GHz by \citet{toothbrush}, see Figure~\ref{fig:tooth_WSRT} right panel. The WSRT observations show that region B3 is polarized at 1.38\,GHz (Figure~\ref{fig:tooth_WSRT} left panel) at a similar level ($\simeq32\pm4\%$) as obtained from the Effelsberg observations at 8.35\,GHz, while regions B2 and B1 have significantly less polarization at 1.38\,GHz compared to the higher frequencies (see Section~\ref{sec:dp}).

	The orientations of the magnetic field lines are changing along the relic, which is unexpected for a continuous shock front. This is best demonstrated by the maps in Stokes U and Q (Figure~\ref{fig:qu}). While the map in Stokes U shows a smooth distribution, two components are clearly separated in the map in Stokes Q that can be identified as B1 and B2+B3. Internal structure of the shock front or Faraday rotation in the foreground are possible reasons (see Section~\ref{sec:rm}).

\subsection{ZwCl\,0008$+$52}

	On the east side of the galaxy cluster ZwCl\,0008$+$52, \cite{zwicky} found a large arc of diffuse radio emission. Based on the location with respect to the cluster center (determined through X-ray observations) and the morphology, they classified the radio source as a radio relic. Furthermore, they did not find any optical counterparts to the diffuse emission, which suggests that the emission is not connected to any cluster galaxy.

	The 8.35\,GHz and 4.85\,GHz images of the relic from our observations are shown in Figure~\ref{fig:zwcl}. We note that the map area of the 8.35\,GHz measurement (left panel) only includes a small region around the eastern relic (indicated by the dashed box in the right panel), whereas the map of the 4.85\,GHz measurement encompasses the entire cluster, including four bright and polarized radio galaxies (named A, B, C and E/F by \citet{zwicky}). The eastern radio relic (RE) can be seen in both images, but only the southern part of the relic is clearly detected at 8.35\,GHz. At 4.85\,GHz, polarized emission is detected along the whole extent of the relic with the B--vectors aligned, while the high-resolution observations of \cite{zwicky} at 1.38\,GHz showed only a few small spots of polarized emission. The western relic cannot be separated from the western radio galaxies in our map at 4.85\,GHz.

	\citet{zwicky} reported a maximum degree of polarization of $\simeq$25\,\% (at 1.38\,GHz) for the eastern relic. Our results show that the relic is polarized with a varying degree of polarization, with a maximum in the southern part of $26 \pm 7\,\%$ at 8.35\,GHz and $22 \pm 4\,\%$ at 4.85\,GHz, while the northern part has a lower degree of polarization of $13 \pm 3\,\%$ at 4.85\,GHz.

	The measurements show that most of the polarization B--vectors are aligned parallel to the major axis of the relic. Nevertheless, further observations are needed to better map the polarization properties. Especially, further polarization observations at 8.35\,GHz are needed with a larger map area and higher sensitivity to detect the western relic and fully trace the eastern relic.


	\begin{figure}[htbp]
		\centering
		\includegraphics[width=9cm, trim=0cm 0cm 0cm 0cm, clip=true]{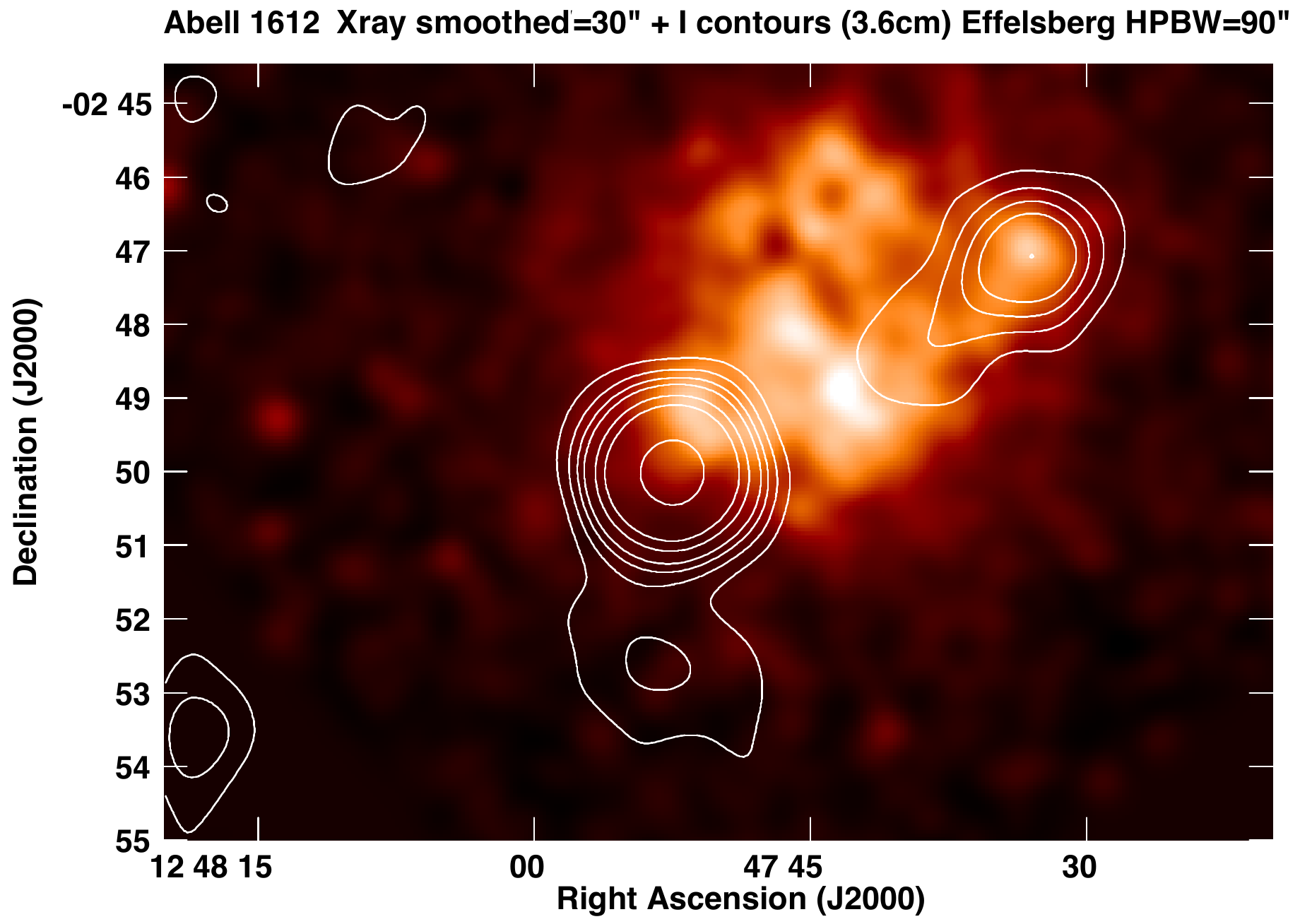}
		\caption{X-ray emission (color) of Abell\,1612 taken with Chandra and smoothed to a beam size of
			$30\arcsec\times30\arcsec$, overlaid with contours of total radio emission at 8.35\,GHz with a beam size
			of $90\arcsec\times90\arcsec$. Contours are drawn at levels of [1, 2, 3, 4, 6, 8, 16]\,$\times\,1.5$\,mJy\,/\,beam.}
		\label{fig:A1612_xray_radio}
	\end{figure}
	
	\begin{figure}
		\centering
		\includegraphics[width=0.45\textwidth]{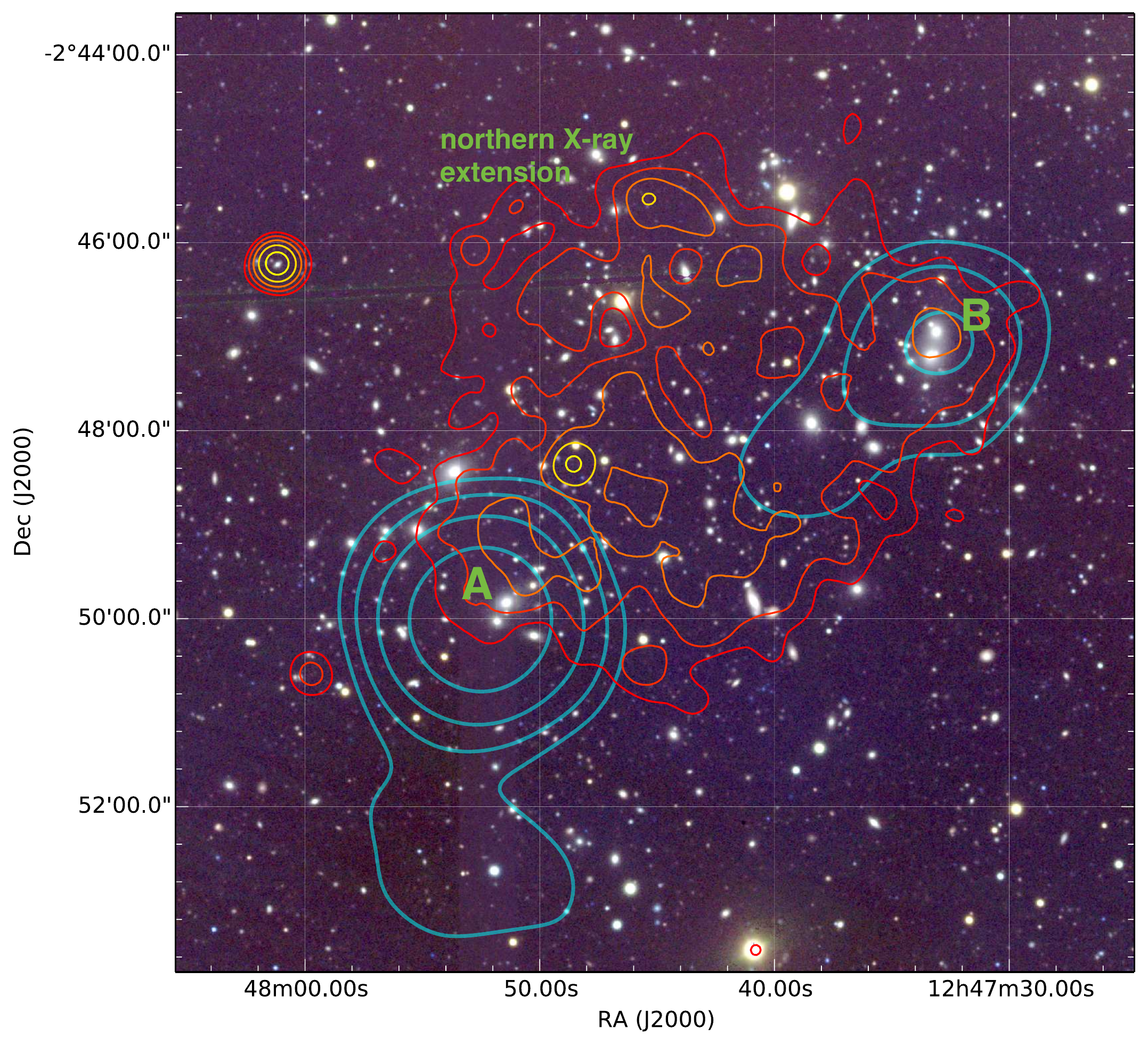}
		\caption{Overlay of a WHT image of Abell\,1612 with the X-ray surface brightness (red--yellow contours) and
the total radio emission at 8.35\,GHz (blue contours).}
		\label{fig:a1612opticaloverlay}
	\end{figure}

\subsection{Abell\,1612}
	\citet{a1612} examined an elongated radio source located in the southern outskirts of the galaxy cluster Abell\,1612. To the north of the extended radio emission there is a bright, tailed radio galaxy. The authors classify the elongated radio source as a candidate for a radio relic because it is apparently not connected to the radio galaxy and they could not find any optical counterpart for the radio source.

	The images shown in Figure~\ref{fig:a1612} present the southern part of the galaxy cluster with a bright radio galaxy in the center (A) and the relic candidate on the southern side (R). Due to the large beam sizes of our observations the detailed structure of the southern relic is smeared out.

    We subtracted bright sources from the Chandra X-ray image and smoothed the image to a beam size of $30\arcsec\times30\arcsec$, to reveal the diffuse X-ray emission, see Figure~\ref{fig:A1612_xray_radio}. The two bright radio sources both have optical counterparts with SDSS spectroscopic redshifts, consistent with being cluster members. The counterparts are massive elliptical galaxies.
	
	The Chandra image displays a clumpy morphology for the distribution of the ICM. The X-ray concentrations near the two radio sources seem to belong to two different subclusters with distinct cD galaxies as judged from optical William Herschel Telescope (WHT) imaging, see Figure~\ref{fig:a1612opticaloverlay}. The structures are separated by about 5\,arcmin, which corresponds to about 900\,kpc. The brighter X-ray emission between the two subclusters originates likely from ram-pressure stripped gas left after the core passage of the two subclusters. Additional X-ray emission is extending to the north which could be from a third merging substructure.
	
	The X-ray surface brightness clearly shows a recent merger of the galaxy cluster. According to the X-ray morphology a large merger shock could be present at the position of the radio relic. Evidently, the relic is  asymmetrically located to the merger axis defined as by the two cD galaxies. Other radio relics are asymmetric with respect to the merger axis as well, see e.g. the relic in the complex merging cluster Abell\,115 \citep{2001A&A...376..803G,2016MNRAS.460L..84B}. The Toothbrush relic is also asymmetric to the main merger axis. \citet{2012MNRAS.425L..76B} suggest a triple merger for the origin. Similarly, the merger in Abell\,1612 apparently comprises several subcomponents.
	
	To determine the global cluster temperature, we extracted a spectrum in a circular region of 1\,Mpc radius, centered on RA=$12^\mathrm{h} 47^\mathrm{m} 43^\mathrm{s}.8$, DEC=$-02\degr\ 48\arcmin\ 00\arcsec$. We fitted this spectrum with {\tt XSPEC} (v12.9), fixing the redshift to z=0.179, the abundance to 0.25, and $N_{\rm{H}} = 1.82\times 10^{20}$\,cm$^{-2}$ \citep[the weighted average $N_{\rm{H}}$ from the Leiden/Argentine/Bonn (LAB) survey,][]{2005A&A...440..775K}. Regions contaminated by compact sources were excluded from the fitting. From this fit we find a temperature of $T=5.5_{-0.3}^{+0.4}$\,keV and a 0.1--2.4\,keV rest-frame luminosity of $1.8 \times 10^{44}$\,erg\,s$^{-1}$.

	In Abell\,1612 no polarization had been detected prior to this work. Our measurements reveal significant polarized emission at 8.35\,GHz. The polarization degree at 8.35\,GHz is $20 \pm 7 \,\%$ at RA\,$\simeq 12^\mathrm{h} 47^\mathrm{m} 56^\mathrm{s}$, DEC\,$\simeq -02\degr\ 52\arcmin$. The polarization degree at 4.85\,GHz cannot be measured with sufficient accuracy because the relic is too close to the central radio galaxy (A). The extensions of polarized emission toward RA\,$\simeq 12^\mathrm{h} 47^\mathrm{m} 55^\mathrm{s}$, DEC\,$\simeq -02\degr\ 54\arcmin$ at 8.35\,GHz and toward RA\,$\simeq 12^\mathrm{h} 47^\mathrm{m} 43^\mathrm{s}$, DEC\,$\simeq -02\degr\ 55\arcmin$ at 4.85\,GHz are at the level of 2.5 times the rms noise and therefore probably not real.

	The high angular resolution images obtained by \citet{a1612} (their Figure 2), together with the fact that these authors did not find any optical counterpart of the elongated radio source in Abell\,1612 and our significant detection of polarization, confirms that this radio source is indeed a radio relic.

\subsection{Integrated flux densities and spectral indices}
\label{sec:flux}

	\begin{table*}
		\centering
		\caption{Flux densities of point sources}
		\begin{tabular}[t!]{lcc}
			\toprule\toprule
			Source		& $S_{\text{I},\,8.35\,\text{GHz}}$ 	& $S_{\text{I},\,4.85\,\text{GHz}}$\\
			Name		& 		(mJy)				& 		(mJy)		\\
			\midrule
			Sausage B		& $4.5\pm1$    & $8\pm2$  \\
			Sausage H	    & $7.5\pm1$  & $13\pm2$ \\
			Toothbrush F	& $3\pm1$    & $5\pm2$  \\
            Abell~1612 A    & $30\pm1$    & $47\pm3$  \\
			\bottomrule
		\end{tabular}
		\label{tab:sources}
	\end{table*}

	\begin{table*}
		\centering
		\caption{Integrated flux densities and fractional polarizations}
		\begin{tabular}[t!]{lcccccccc}
			\toprule\toprule
			Relic		& $S_{\text{I},\,8.35\,\text{GHz}}$	&$S_{\text{PI},\,8.35\,\text{GHz}}$	&  $p_{8.35\,\text{GHz}}$	&$S_{\text{I},\,4.85\,\text{GHz}}$&$S_{\text{PI},\,4.85\,\text{GHz}}$	 &$p_{4.85\,\text{GHz}}$ &$\alpha$&$\cal{M}$\\
			Name& 		(mJy)				& 		(mJy)		&(\%)		&(mJy) & (mJy)	&(\%)&& \\
			(1)      	&(2)							&(3)             			&(4)		&(5)&(6)&(7)	&(8)&(9)	\\
			\midrule
			Sausage		& $17\pm5$     & $5.0\pm0.5$    & $29\pm9$    & $32\pm8$	     & $11.4\pm1.0$ & $36\pm10$	&$0.90\pm0.04$	&$-$\\
			Toothbrush	& $59\pm7$     & $13.1\pm0.8$	& $22\pm3$     & $68\pm5$	     & $10.2\pm0.3$ & $15\pm1$	&$1.00\pm0.04$	&$-$\\
			ZwCl\,0008$+$52	& $5.0\pm1.5$  & $1.1\pm0.4$	& $22\pm10$	& $18.3\pm3.6$	 & $2.3\pm0.4$  & $13\pm3$	&$1.44\pm0.04$		&$2.35\pm0.1$ \\
			Abell\,1612		& $5.6\pm2.2$  & $0.7\pm0.2$	& $13\pm6$	& $9.9\pm2.1$	 & $0.5\pm0.2$  & $5\pm2$ 	&$1.39\pm0.13$		&$2.47\pm0.3$\\
			\bottomrule
		\end{tabular}
		\tablefoot{(1) Source name; (2) total flux density at 8.35\,GHz; (3) polarized flux density at 8.35\,GHz; (4) polarization fraction at 8.35\,GHz; (5) total flux density at 4.85\,GHz; (6) polarized flux density at 4.85\,GHz; (7) polarization fraction at 4.85\,GHz; (8) integrated spectral index; (9) Mach number derived from the spectral index of the integrated flux densities.}
		\label{tab:fluxes}
	\end{table*}

	The integrated flux densities in total and polarized intensity were obtained by integrating within $1\sigma$ (Table \ref{tab:fluxes}). Thereby for all relics and both frequencies, we defined different integration areas around the relic down to the noise level (see Tables \ref{tab:observation3cm} and \ref{tab:observation6cm}).

	To correct the baselevel of each map in Stokes I, U, and Q, the mean intensity of each map was measured by selecting at least five boxes (each containing about one beam size) in regions where no emission from sources is detected. The average of these values was subtracted from the final map. Since the sizes of some maps are too small to find a sufficiently large number of regions without signal, the uncertainties in the baselevel subtraction are quite large and affect the integrated flux densities. Furthermore, for the same reason the baselevels may not be homogeneous (large-scale fluctuations) in the maps which leaves artifacts in total and polarized intensity.

	The point sources external to the Sausage relic, named B and H by \citet{CIZA}, and the source F external to the Toothbrush relic (see Figure~\ref{fig:sausage} left and Figure~\ref{fig:tooth} left) were subtracted from maps of the total intensity at 8.35\,GHz before performing the integration. As these three sources are not resolved at 4.85\,GHz, we extrapolated their flux densities from 8.35\,GHz by assuming a spectral index of 1.0. The resulting additional error in the integrated flux density at 4.85\,GHz is much smaller than the error due to the baselevel uncertainty (see above). For Abell\,1612, the bright source north of the relic (``A'') was subtracted from the maps before performing the integration. The flux densities of the subtracted point sources are given in Table~\ref{tab:sources}. For ZwCl\,0008$+$52, no point sources had to be subtracted.


	The uncertainty of the flux density measurements is dominated by the uncertainty in the baselevel of a map. The uncertainties due to rms noise in a map and the uncertainties in the procedure of source subtraction are much smaller and can be neglected. The absolute calibration error of about 5\% affects all flux densities in the same way and is disregarded in the following.

	To obtain the uncertainty $\Delta_s$ of a flux density measurement, we determined the standard deviation $\sigma_s$ of the mean intensities of the boxes where the baselevel was measured. As this uncertainty systematically affects the flux density of the source, we multiplied $\sigma_s$ with the number of beams $N_{\text{beam}}$ within the integration area:
	\begin{equation}
	\label{eq:deltaS}
		\centering
		\Delta_s
		=
		\sigma_s \cdot N_{\text{beam}}
	\end{equation}

	This equation is applicable if the scale of the fluctuations in the baselevel is larger than the size of the integration area, while smaller-scale fluctuations would cancel out in the integration process.
	
\begin{figure*}
	\centering
	   \includegraphics[width=9cm, trim=4cm 1cm 5cm 2cm, clip=true]{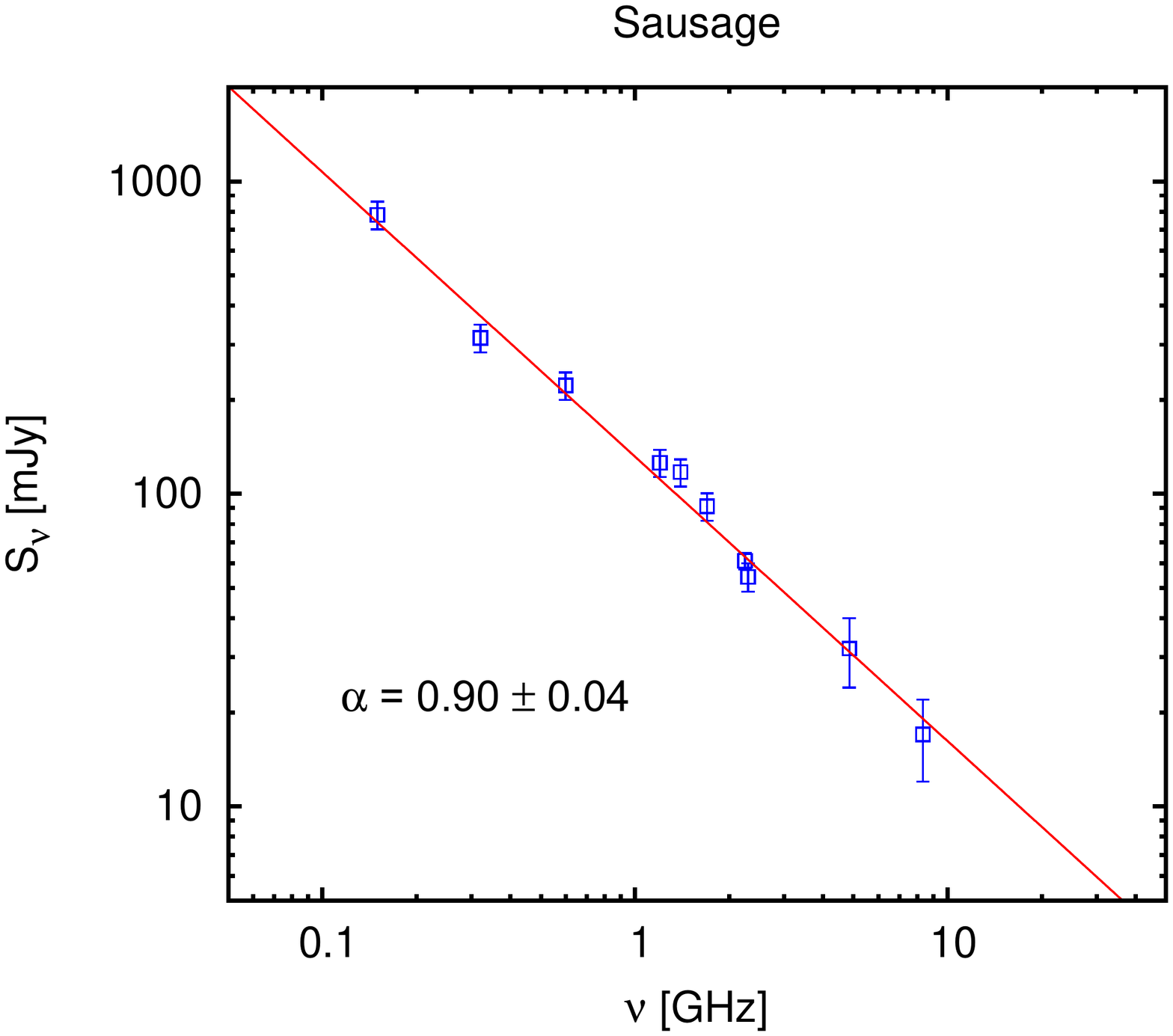} \includegraphics[width=9cm, trim=4cm 1cm 5cm 2cm, clip=true]{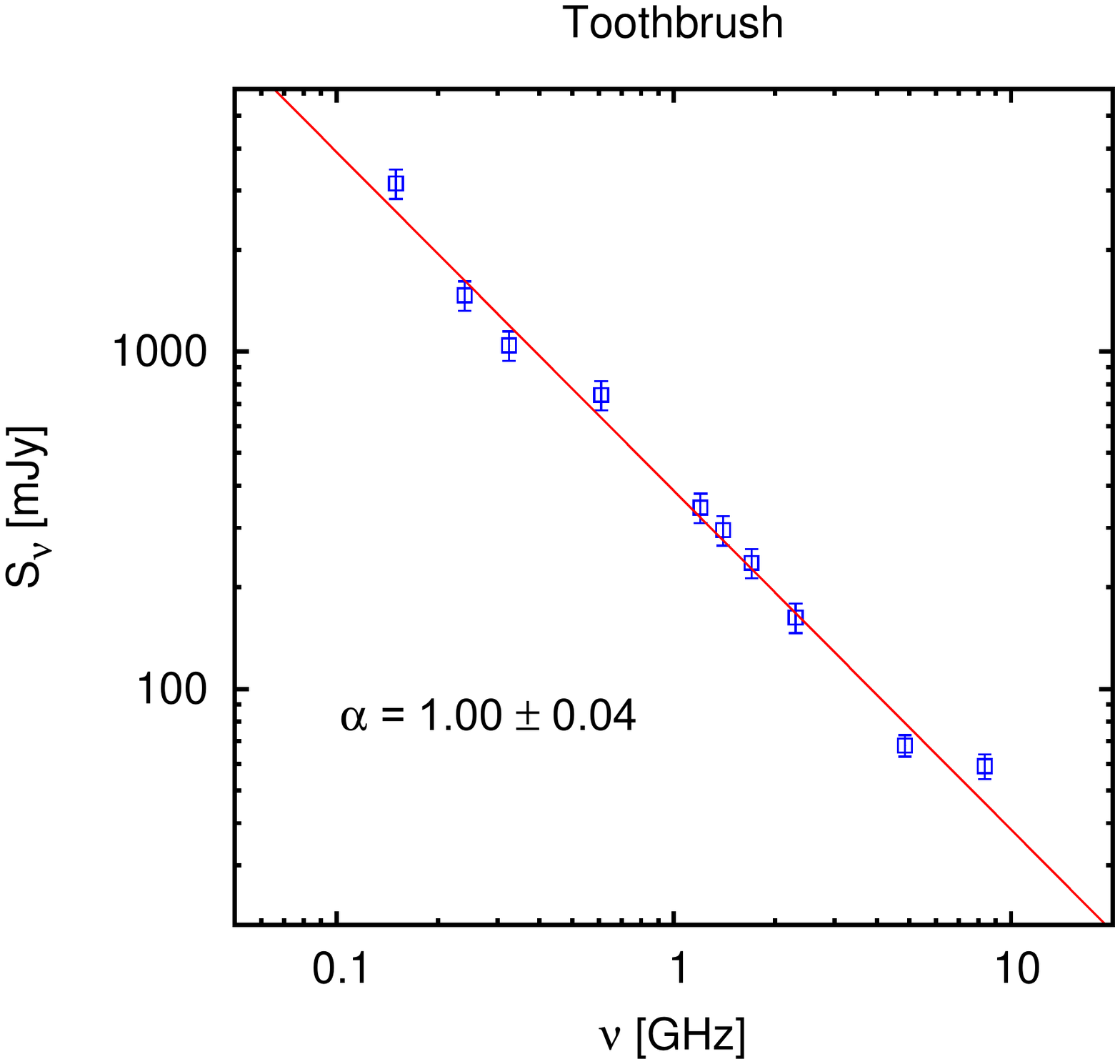}
   	   \includegraphics[width=9cm, trim=4cm 1cm 5cm 2cm, clip=true]{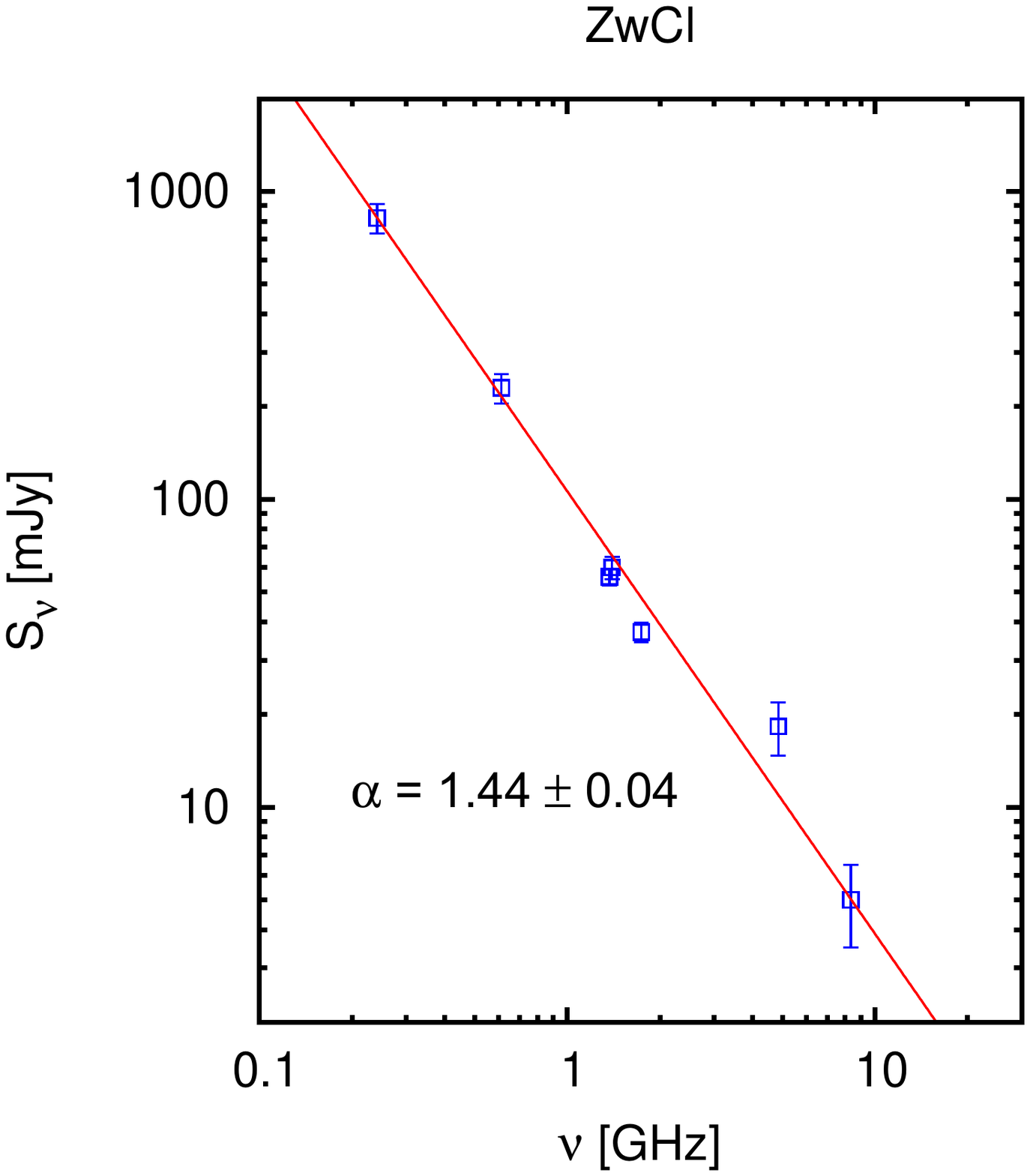} \includegraphics[width=9cm, trim=4cm 1cm 5cm 2cm, clip=true]{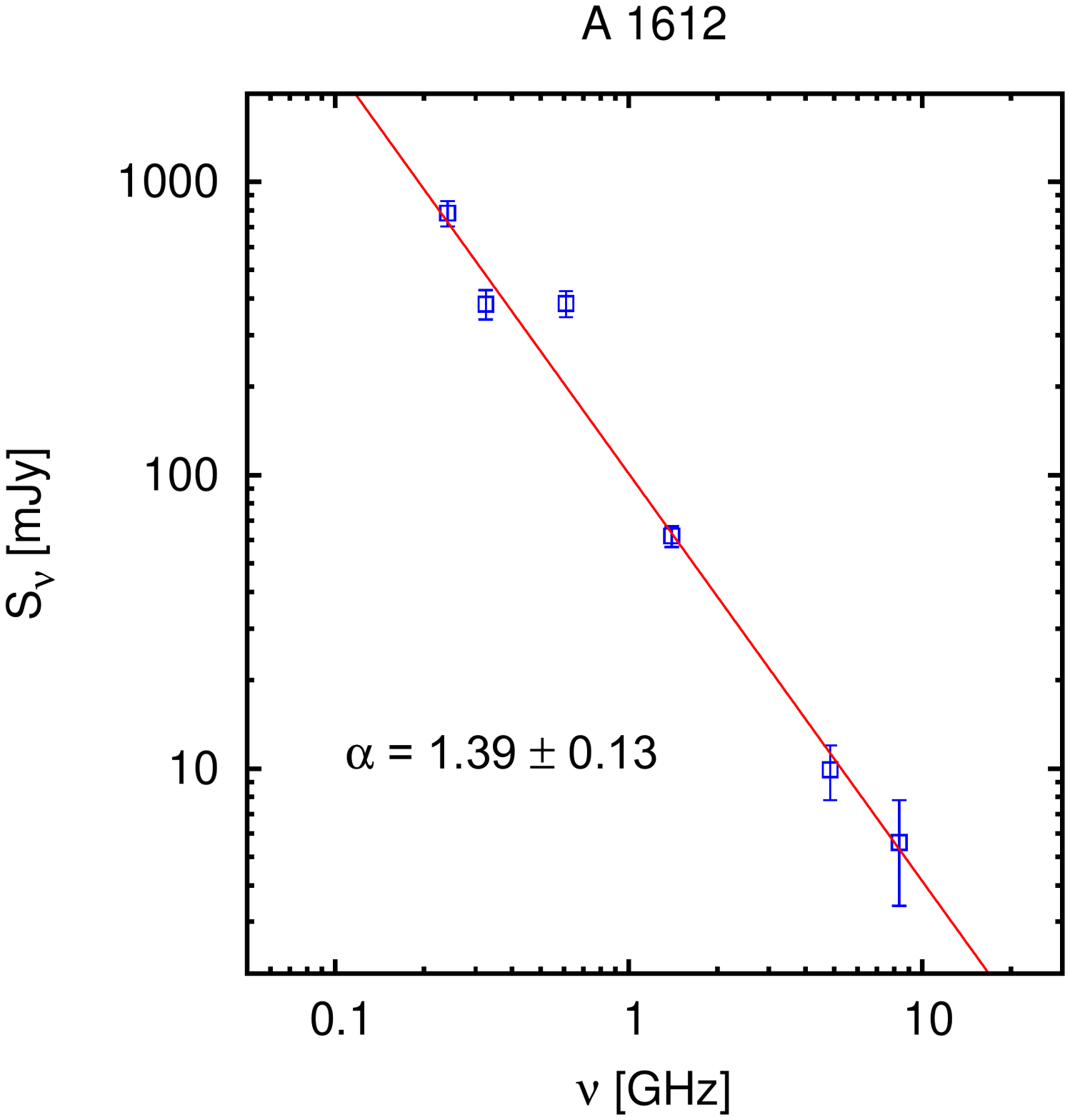}
	     \caption{Integrated total intensity spectra of the relics considered in this paper.
	     \textit{Top Left:} Spectrum of the Sausage radio relic. The flux densities at frequencies lower than 4.85\,GHz are taken from Table~3 of \citet{2016MNRAS.455.2402S}.
	     \textit{Top Right:} Spectrum of the Toothbrush radio relic. The flux densities at frequencies lower than 4.85\,GHz are taken from Table~3 of \citet{2016MNRAS.455.2402S}.
     	     \textit{Bottom Left:} Spectrum of the radio relic in ZwCl\,0008$+$52. The flux densities at frequencies lower than 4.85\,GHz are taken from \citet{zwicky}.
     	     \textit{Bottom Right:} Spectrum of the radio relic in Abell\,1612. The flux densities at frequencies lower than 4.85\,GHz are taken from \citet{a1612}.}
	     \label{fig:spectra}
	\end{figure*}

	The plotted spectra for all relics considered in this paper are shown in Figure \ref{fig:spectra}. We decided to use the flux densities at frequencies below 4.85\,GHz from Table~3 of \citet{2016MNRAS.455.2402S} which were obtained from images from uv data beyond 800\,$\lambda$, corresponding to angular scales of smaller than about $3\arcmin$. The reasons are the following: (1) The Sausage relic shows diffuse extended emission at 153\,MHz but not at higher frequencies \citep{CIZA}, indicating that this emission has a steep spectrum. Hence little additional emission is expected when using images based on uv data smaller than 800\,$\lambda$. However, the flux densities in Table~4 of \citet{2016MNRAS.455.2402S} are higher than those in Table~3 of \citet{2016MNRAS.455.2402S} by factors of 1.4--2.1. (2) The flux density of the Toothbrush relic at 1.4\,GHz in Table~3 agrees well with that at 1.5\,GHz by \citet{2016ApJ...818..204V}, based on JVLA D-array data that hardly suffer from missing large-scale emission.

For all four relics we find flux densities which are in accordance with power law distributions at frequencies below 4.85\,GHz.
We do not find any evidence for significant spectral steepening at frequencies below 8.35\,GHz in any of the four relics. This also rules out a significant B-field variation within the CRE cooling time. The steepening found by \citet{2016MNRAS.455.2402S} for the Sausage and Toothbrush relics are based on two values at higher frequencies which however have large errors and may suffer from the SZ effect \citep{2016A&A...591A.142B}. We note that the integrated flux densities of the Sausage and Toothbrush relics given in Table~4 of \citet{2016MNRAS.455.2402S} are based on the same maps as in this work, but are slightly different because the values in our Table~\ref{tab:fluxes} are obtained with somewhat different integration areas and an additional source subtraction in the case of the Toothbrush relic.
	
	The flux density of the Toothbrush relic at 4.85\,GHz in Table~\ref{tab:fluxes} is possibly too low compared to 8.35\,GHz and the values at lower frequencies (Figure \ref{fig:spectra}). The probable reason is the insufficient size and low quality of this map. Interestingly, our results indicate that the radio spectrum of the Toothbrush runs up to 8.35\,GHz with a constant spectral index of about 1.0. In X-rays, only a moderately strong shock has been reported, with a Mach number of about 1.5 \citep{2015PASJ...67..113I,2016ApJ...818..204V}.
	

	According to diffusive shock acceleration (DSA) theory in the test particle regime, particles get accelerated at a shock front to relativistic energies showing a power-law distribution in particle energy \citep{1987PhR.154.1B}. The strength of the shock, described by the Mach number or the compression ratio, determines the exponent, $s_\text{inj}$, of the energy spectrum. Radio relics are straightforwardly explained with DSA: The merger shock injects CREs into the ICM. While the shock front propagates further out, the CREs cool. If the shock properties do not change on a time scale comparable to the cooling time of the CREs that generate the observed synchrotron emission, the CRE energy distribution is in steady-state, determined by continuous injection and synchrotron and inverse Compton losses. The resulting overall CRE energy spectrum is again a power law ($N(E)\propto E^{-s}$) with $s = s_\text{inj} + 1$ \citep[e.g.][]{1999ApJ...520..529S}. The overall radio spectrum is therefore a power-law as well. The spectral index $\alpha$ of the spectrum of flux densities integrated over the relic (see Table \ref{tab:fluxes}) is related to the Mach number $\cal{M}$ according to:

	 \begin{equation}
		\centering
		\cal{M}
		=
		\sqrt{\frac{\alpha +1}{\alpha -1}}~.
		\label{eq:mach}
	\end{equation}

	Evidently, in the DSA scenario the radio spectrum is a power law with spectral index steeper than 1. A bent spectrum is obtained, for instance in case of re-acceleration of a pre-existing mildly relativistic electron population or with a variable downstream magnetic field strength \citep{2016MNRAS.462.2014D}.
	We found spectral indices of $0.90 \pm 0.04$ and $1.00 \pm 0.04$, for the Sausage and the Toothbrush, respectively, indicating that models describing the origin of relics have to include effects beyond the assumptions made above. For instance, if the Mach number and the injection electron spectrum changes on time scales comparable to the cooling time, the steady-state assumption is violated and the resulting spectrum may deviate from Equation~\ref{eq:mach}. Measuring the spectral index close to the shock front with high angular resolution provides a more reliable estimate for the injection spectrum.
	
	We note that for the Toothbrush only a moderately strong shock has been reported from X-ray observations, with a Mach number of about 1.5 \citep{2015PASJ...67..113I,2016ApJ...818..204V}. According to Equation~\ref{eq:mach} this corresponds to a spectral index of 2.6, evidently not in accord with the observations. It is a matter of an ongoing debate if X-ray observations may tend to indicate too low Mach numbers or if the acceleration mechanism is far more complex than DSA of thermal or mildly relativistic electrons. 	

	For the eastern relic in ZwCl\,0008 we confirm a steep spectral index as reported by \citet{zwicky}. The flux measurements of the relic in Abell\,1612 are rather uncertain due to the proximity of the tailed radio galaxy. In combination with flux density measured at 1.38\,GHz our measurements indicate, however, a spectral index of about 1.4. For both relics the spectral index is rather steep compared to relics associated with merger shocks \citep{2012A&ARv..20...54F}. According to DSA in the test-particle regime, the spectral index corresponds to a Mach number of about 2.4.

\subsection{Magnetic field strengths}
\label{sec:mf}

	\begin{table}
	\caption{Equipartition magnetic field strengths (assuming a proton-to-electron ratio of $K=100$) }
	  \begin{tabular}[t]{lcc}
	  	\toprule\toprule
	  	Name (component)             & $B_\mathrm{tot}$ & $B_\mathrm{an}$\\
	  	                             & ($\muG$)         & ($\muG$)       \\
	  	(1)                          & (2)              & (3)            \\
	  	\midrule
	  	Sausage (east)               & 2.4              & 2.0            \\
	  	Sausage (middle)             & 2.5              & 2.0            \\
	  	Sausage (west)               & 2.4              & 1.8            \\
	  	Toothbrush (east, comp B3)   & 2.6              & 1.6            \\
	  	Toothbrush (middle, comp B2) & 3.0              & 2.5            \\
	  	Toothbrush (west, comp B1)   & 4.2              & 2.0            \\
	  	ZwCl\,0008$+$52              & 2.4              & 1.5            \\
	  	Abell\,1612                  & 2.7              & 1.5            \\
	  	\bottomrule
	  \end{tabular}
	   \tablefoot{(1) Source name and component; (2) total magnetic field strength;
	   (3) strength of the anisotropic field. The field strengths have been corrected for inclination effects.}
	   \label{tab:B}
	\end{table}

 	\begin{figure}
 		\centering
 		\includegraphics[width=0.45\textwidth]{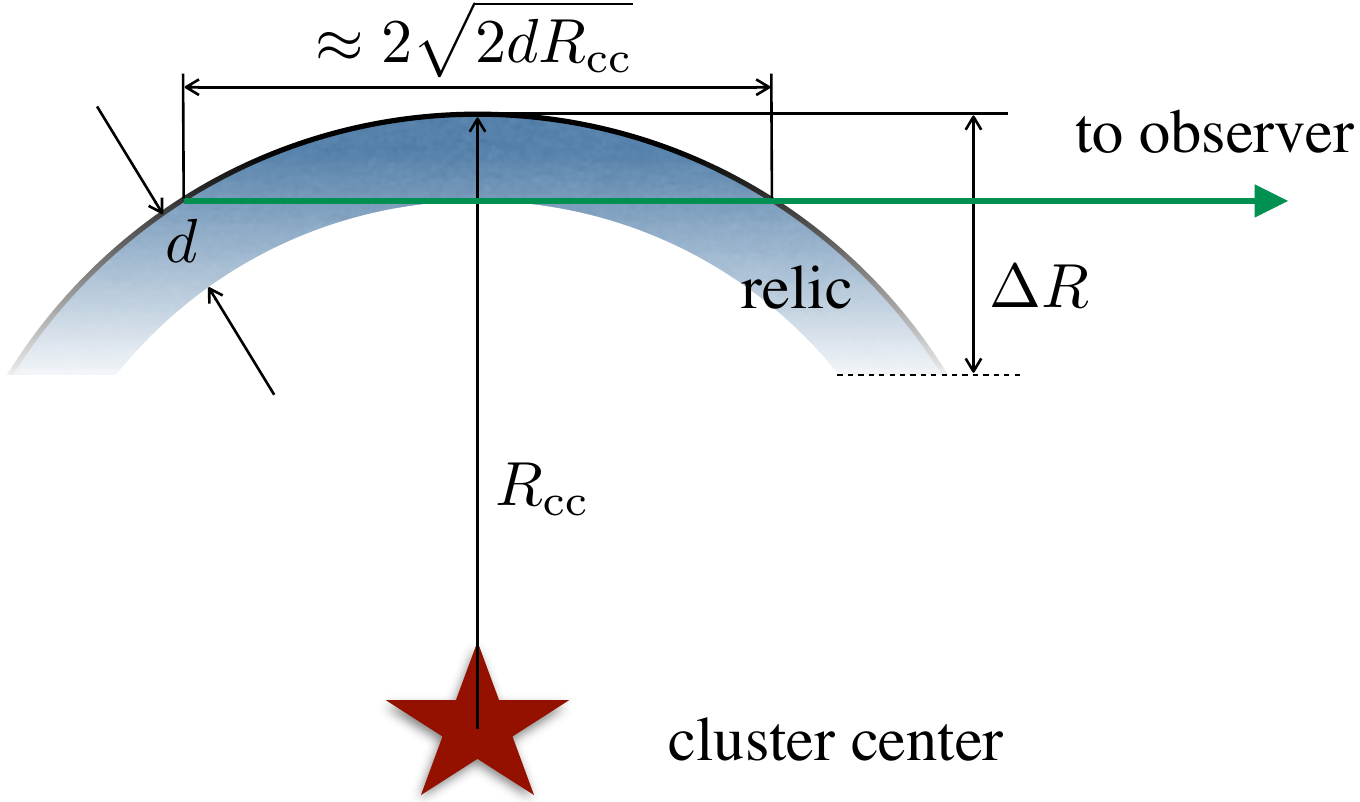}
 		\caption{Sketch for the path length through the emitting volume of a relic.
			 The width of the downstream volume is given by the cooling time of the electrons. The green line indicates the line of sight with the maximum path length through the emitting volume. $d$ is the intrinsic width and $\Delta R$ the observable width of the relic (see Table~\ref{tab:parameters}). For $d \ll R_\text{cc}$ the maximum path length can be approximated by $2\sqrt{2\,d\,R_\text{cc}}$. }
 		\label{fig:sketch_path_length}
 	\end{figure}

	The synchrotron emission from radio relics gives evidence that the intra-cluster medium in the downstream region of the shocks is significantly magnetized. About 50 radio relics have been detected so far, however, very little is known about the strength and structure of magnetic fields in relics, see \citet{2004IJMPD..13.1549G} for a summary. The radio intensity alone does not provide information on the field strength without additional assumptions, since the density of relativistic electrons is not known. The most promising approach to measure the latter is to detect the inverse Compton (IC) signature of relativistic electrons in the hard X-ray band. So far, only upper limits for the IC emission have been obtained. They provide lower limits for field strength in relics when combining with the radio luminosity. For two relics, namely Abell\,3667 and the Toothbrush relic, \citet{2010ApJ...715.1143F} and \citet{2015PASJ...67..113I} concluded that the field strengths in the relic regions are larger than 3\,$\mu$G and 1.6\,$\mu$G, respectively.
	
	One possibility to break the degeneracy between relativistic electron density and magnetic field strength is to assume energy equipartition between magnetic fields and cosmic rays (CRs). This has lead to reasonable estimates for the total magnetic field strengths \citep{2005AN....326..414B}.

\citet{2001AJ....122.1172S} observed radio-bright relics in four galaxy clusters and found a total field strengths of 7--14\,$\mu$G from the minimum energy assumption, which gives similar results as the equipartition assumption, and adopting a proton-to-electron ratio of $K=1$. However, all four relics are classified as ``radio phoenix'', i.e. they are believed to originate from radio galaxies, possibly re-energized by shock compression \citep{2001A&A...366...26E}. Therefore, these objects likely comprise a stronger magnetic field than the relics associated with large merger shocks.

	Another approach to measure the magnetic field strength in galaxy clusters is the analysis of the Faraday rotation of polarized sources in the background or embedded in the ICM. The rotation of the polarization direction measures the average line-of-sight component of the magnetic field weighted with the thermal electron density. \citet{2013MNRAS.433.3208B} studied the rotation measures of background sources close to the radio relic in the Coma cluster. They derived a strength of the line-of-sight field	of $\ge$\,2\,$\mu$G, basically in agreement with the lower limits and estimates as discussed above.
	
	We apply the revised energy equipartition formula \citep{2005AN....326..414B} to estimate the magnetic field strength. Using this method, the CR spectrum is integrated down to $E\,=\,0$, assuming a spectral break at the rest mass of the CR protons at 938 MeV, as expected for a DSA spectrum that is a power law in momentum space. If however the CR energy is dominated by leptons, the integration should be performed over the electron spectrum. As the low-energy electrons below a few 100 MeV suffer strongly from energy losses, a spectral break at 936 MeV still seems like a useful assumption and should yield reasonable equipartition values.

The strength of the total magnetic field $B_{\mathrm{tot},\perp}$ in the sky plane is derived from the total synchrotron intensity $I_\mathrm{sync}$ via:
	\begin{equation}
		 B_{\mathrm{tot},\perp} \propto [I_\mathrm{sync}/(L \, (K+1))]^{\,\,\,1/(3+\alpha_{\text{inj}})} \, .
		\label{eq:equi}
	\end{equation}
	To derive field strengths, assumptions for the proton-to-electron ratio $K$ of the number densities of CR particles in the GeV range, the path length $L$, and the synchrotron spectral index $\alpha_{\text{inj}}$ at the location where the CRs are accelerated have to be made.

The radio emission probes only the CR electron content. CR protons in galaxy clusters have not been detected, so that we can only speculate how much energy is stored in CR protons. Diffusive particle acceleration typically predicts $K$ of the order of $K=100$ \citep{1978MNRAS.182..443B}. However, such a high proton-to-electron ratio may violate the upper limits from observations with the FERMI telescope \citep{2015MNRAS.451.2198V}.
	
	It has been recently suggested that at low Mach number shock front electrons might be scattered back and forth at the shock, as assumed for a Fermi~I-type process, and gain energy via shock drift acceleration (SDA) \citep{2014ApJ...794..153G}. Such a process accelerates only electrons effectively. If electrons at merger shock fronts are accelerated by SDA, the efficiency for proton acceleration might be negligible since the Mach numbers of merger shocks are low, $\cal{M} \lesssim$\,4, and DSA would not be efficient enough to accelerate protons, implying a proton-to-electron ratio close to $K=0$. To back scatter electrons in the upstream region, the large-scale guiding magnetic field needs to be disturbed. As shown by \citet{2014ApJ...797...47G}, the Firehose instability may sufficiently twist the field lines, provided that the upstream plasma has a low magnetization, i.e. the ratio of thermal to magnetic energy densities has to be large, $\beta > 20$.

	In addition to $K$, the synchrotron spectral index $\alpha$ and the path length $L$ through the synchrotron-emitting medium need to be known. The integrated $\alpha$ values in Table \ref{tab:fluxes} represent the spectrum of cosmic-ray electrons which is steepened by synchrotron losses. As equipartition holds for the freshly injected particles, we assume injection values of $\alpha_{\text{inj}}=0.7$ for the Sausage relic \citep{CIZA} and the Toothbrush relic \citep{toothbrush}. For ZwCl\,0008$+$52 and Abell\,1612 we correct $\alpha$ for synchrotron losses by 0.5 which gives $\alpha_{\text{inj}}=0.9$.

The intrinsic width of a radio relic, i.e. the downstream extent of the emitting volume, can be estimated from the downstream velocity and the magnetic field strength. The velocity of the downstream plasma with respect to the shock is basically given by the downstream specific energy density \citep{2007MNRAS.375...77H}; for a plasma with a temperature of 7\,keV the velocity amounts to about 1400\,km/s. Adopting that velocity, 66\,\% of the emission at about 5\,GHz originates from a volume within a distance of about 10\,kpc to the shock front (Hoeft et al., in prep.). In observations, relics are more extended into the downstream direction, see Table~\ref{tab:parameters} (Column~8). Smoothing due to the resolution of the observation and projection effects (since the shock fronts may not be aligned with the line of sight) define the observed width.

	Simulations from \citet{2011MNRAS.418..230V} show that a calotte of a sphere is a good approximation for the shape of a relic. Figure~\ref{fig:sketch_path_length} illustrates an estimate for the path length through the emitting volume. We used this simple geometry to calculate the average path length $L$ through the relic from the distance to the cluster center, $R_\text{cc}$, and the intrinsic width, $d$. We assume that the average path length for all lines of sight through the emitting volume of the relic is $1/\sqrt{2}$ of maximum path length, hence
	\begin{equation}
		L \approx 2 \sqrt{ d  \, R_\text{cc} } .
	\end{equation}
	Adopting for the intrinsic width 10\,kpc and for the distance to cluster center 1.0--1.5\,Mpc, we obtain for the average path length through the emitting volume in the four relics 200--250\,kpc.
	
	Eq.~\ref{eq:equi} yields the field component $B_{\mathrm{tot},\perp}$ perpendicular to the line of sight, giving rise to synchrotron emission. The shock front generating the relic amplifies the field components in the plane of the shock front that is seen primarily edge-on. Hence, the field components along the relic and along the line of sight are enhanced. As synchrotron emission emerges only from the component in the sky plane, only emission from the compressed field along the relic is observed. To correct for the invisible field components along the line of sight, we require an average angle between all possible orientations of the field in the plane of the shock front and the line of sight. We assume this angle is $45\degr$.

	Assuming $L=200$\,kpc and $K=100$ yields total field strengths in the range 2.4--4.2\,$\mu$G (Table~\ref{tab:B}). The value for the Sausage relic is consistent with the estimate of 5--7\,$\mu$G derived from the width of the relic by \citet{2010Sci...330..347V}. For $K=0$ all values are smaller by the factor $101^{\, -1/(3+\alpha_\mathrm{inj})}$ (see Eq.~\ref{eq:equi}) and become 0.7--1.2\,$\mu$G.

	The largest uncertainties of $B_\mathrm{tot}$ emerge from the values assumed for the path length $L$ and the ratio $K$. Changing one of these values by a factor $a$ changes the field strength by a factor of $a^{\,- 1/(3+\alpha)}$ (Eq.~\ref{eq:equi}). Due to the simplified geometrical model of the relic shape, we estimate an uncertainty of $a\le3$, resulting in an error in $B_\mathrm{tot}$ of $\lesssim$\,30\%.

	The component of the magnetic field that is aligned along the shock front, called the anisotropic field $B_\mathrm{an}$ (see Section~\ref{sec:mach}), is computed from the degree of linear polarization and is also listed in Table~\ref{tab:B}. Due to the limited resolution of our observations, the strengths of the anisotropic field in Table~\ref{tab:B} are lower limits and may in fact be very similar to the total field strengths.
	
	Adopting a typical electron density in the cluster periphery of $10^{-4}\:\rm cm^{-3}$ \citep{2008A&A...487..431C} and a temperature of 8\,keV as found for the Toothbrush \citep{2016ApJ...818..204V}, this results in a plasma--$\beta$ (the ratio between the energy densities of thermal gas and magnetic fields) in the range 1.3 to 14, where the higher value corresponds to the lower field strength. It is not clear if energy equipartition or minimum energy arguments apply for the field in the downstream plasma of merger shocks. However, the considerations presented above permit a scenario in which only electrons are accelerated, i.e. $K=0$, so that the lower field estimates apply, $B\approx 1\:\rm \mu G$. This value reflects the strength in the downstream regime, where the field strength is presumably higher due to compression at the shock front, or CR-induced instabilities, and/or turbulent amplification. In a much weaker upstream field ($\beta \gtrsim 10$), Firehose instabilities can generate upstream waves, which allow multiple energy gains of electrons enabling a Fermi~I-type process for electron acceleration at a low Mach number shock. A significantly stronger magnetic field would not permit such a process.

	On the other hand, the non-detection of inverse Compton emission from the Toothbrush relic yields a lower limit for the field strength in downstream region of 1.6\,$\mu$G \citep{2015PASJ...67..113I}. Furthermore, \citet{2016A&A...591A.142B} estimated a total field strength of the Sausage relic of about 2.5\,$\mu$G from the flux ratio between the Sunyaev-Zel'dovich (SZ) and the synchrotron signals and with an ad-hoc assumption for the electron acceleration efficiency. Field strengths of a few $\mu$G in the downstream region correspond to a plasma--$\beta$ about unity. Hence, these results favor the strong-field scenario where both protons and electrons are accelerated in the shock	front.


\subsection{Mach numbers derived from the degree of polarization}
\label{sec:mach}

	\begin{figure*}
		\centering
	   \includegraphics[width=8.5cm, trim=1cm 7cm 1cm 2cm, clip=true]{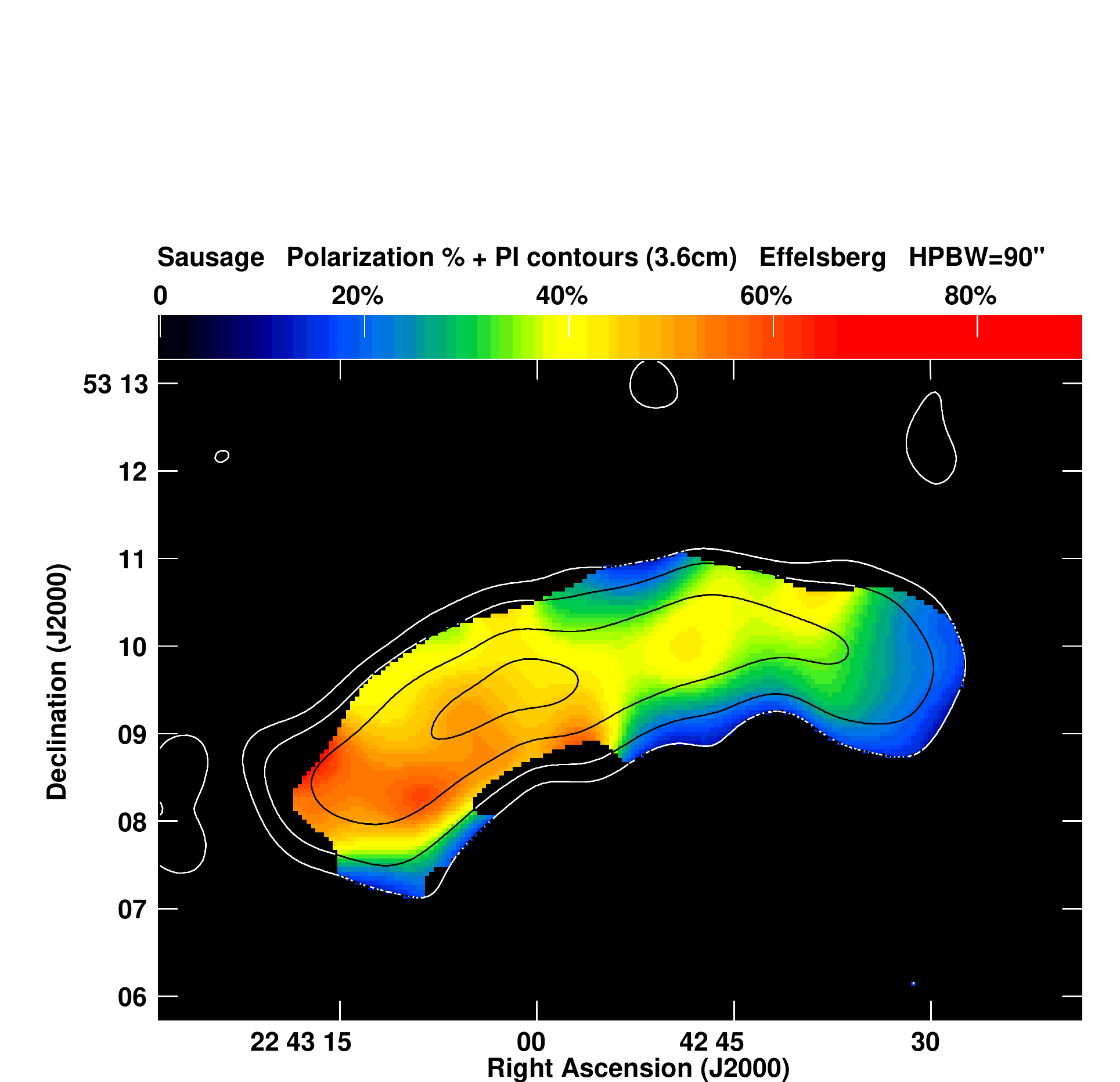}
	   \hspace*{-1cm}\hspace*{10mm}
	   \includegraphics[width=9cm, trim=1cm 7cm 1cm 2cm, clip=true]{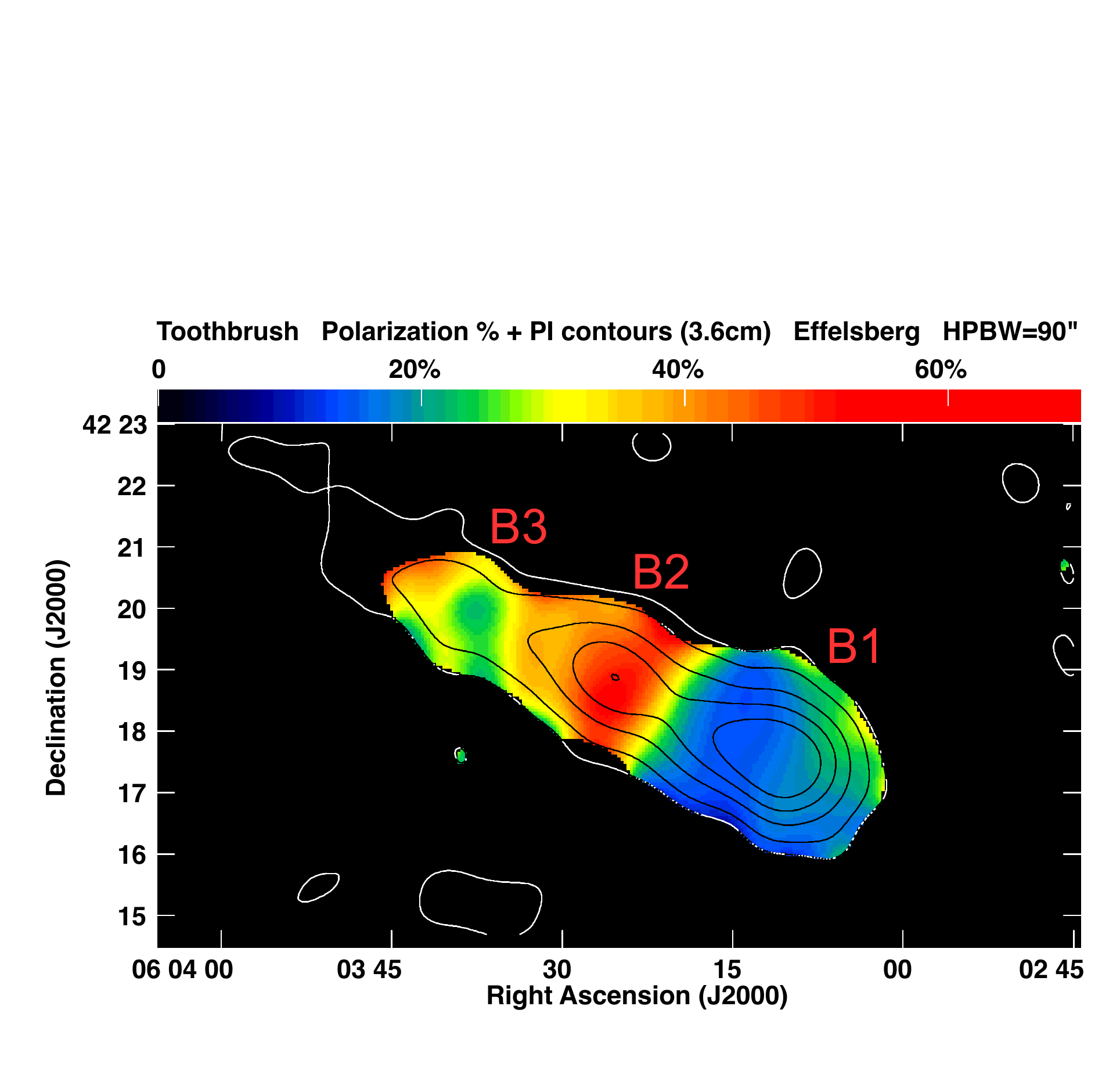}
	   \caption{Degree of polarization (color) of the Sausage relic (\textit{left}) and the Toothbrush
	     relic (\textit{right}) at 8.35\,GHz with $90\arcsec \times 90\arcsec$ resolution. The contours show the
polarized intensity at [1, 2, 4, 6]\,$\times\,0.21$\,mJy\,/\,beam for the Sausage relic (\textit{left}) and at
	     [1, 2, 4, 6, 8]\,$\times\,0.39$\,mJy\,/\,beam for the Toothbrush relic (\textit{right}).}
	     \label{fig:percpol}
	\end{figure*}

	\begin{figure}
		\centering
		\includegraphics[width=0.4\textwidth]{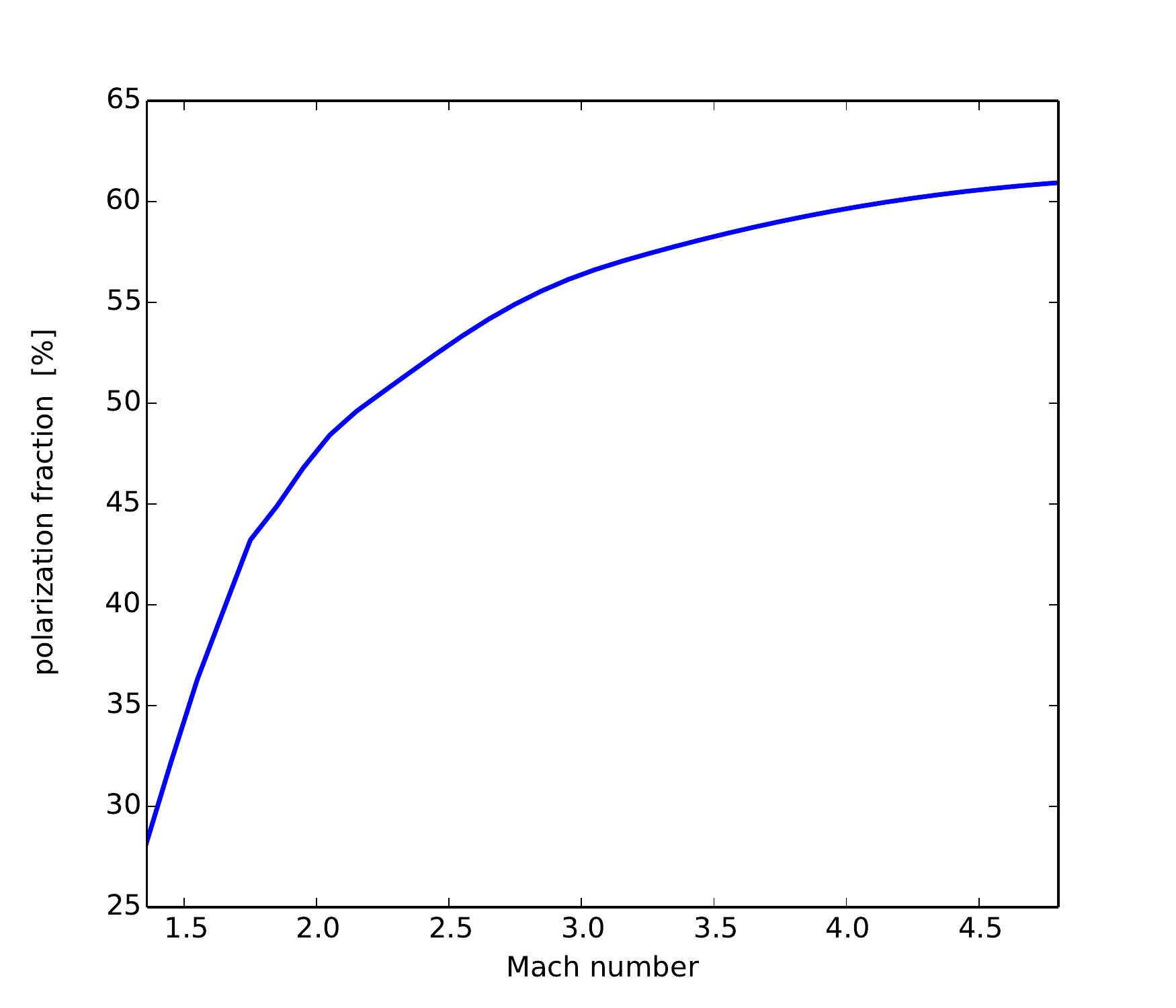}
		\caption{The degree of polarization as a function of the Mach number of the shock. The polarization is derived
			assuming a weak upstream magnetic field that is isotropically tangled on small scales and compressed at the shock front (Hoeft et al., in prep).}
		\label{fig:pol_vs_mach}
	\end{figure}

	\citet{ensslin} suggested that shock fronts in the ICM induced by cluster mergers are the origin of radio relics. In the proposed scenario electrons are accelerated at the shock front via diffusive shock acceleration. While the shock front propagates further, the relativistic electrons cool via synchrotron and inverse Compton emission. Furthermore, \citet{ensslin} argue that the relics should be polarized due to compression of turbulent or tangled magnetic fields at the shock front, see also \citet{1980MNRAS.193..439L}. This model can be briefly summarized as follows: The medium ahead of the shock fronts is permeated by magnetic fields that are turbulent or tangled on scales many orders of magnitude smaller than the length scale corresponding to the telescope beam. The distribution of magnetic field vectors in the upstream medium is isotropic. Shock compression entails that the magnetic field components parallel to the front are stretched and amplified, while the perpendicular components remain constant. As a result, the magnetic field behind the shock front is still randomly tangled, but the distribution of field orientations is not isotropic anymore (also called an anisotropic field). It is stretched parallel to the plane of the shock. This causes the synchrotron emission to be polarized, depending on the degree of anisotropy caused by the shock front.
	
	In a numerical model, described in detail in a forthcoming paper (Hoeft et al., in prep), we study the Mach number dependence of the fractional polarization (see Figure \ref{fig:pol_vs_mach}). A shock with a Mach number of 2.0 already causes a fractional polarization of about 48\,\%. However, it should be kept in mind that this is computed for relics seen perfectly edge-on and assuming no further depolarization, for instance due to variations in the orientation of the shock surface. Interestingly, the relics in ZwCl\,0008 and Abell\,1612 shows a steeper spectral index in the radio spectrum and also a lower degree of polarization, both indicating a low Mach number.

	\citet{2016ApJ...818..204V} studied the Toothbrush relic in detail. They found that the spectral index at the northern edge varies in the range $\alpha_{150}^{610} \approx 0.7 - 0.8$. The regions of brighter emission along the edge correlate with the regions showing a flatter spectrum. Due to the high resolution in the images, the measured spectral index at the edge should be close to the injection spectrum. Adopting DSA implies Mach numbers in the range ${\cal M} \approx 2.8 - 3.3$. Hence, the intrinsic polarization due to a compressed magnetic field should be about 50\% to 55\%. Actually, our observation shows for 8.35\,GHz a polarization of about 45\% for the region B2. According to Figure \ref{fig:pol_vs_mach}, this implies that the minimum Mach number in that region is 2.0. However, even if the Mach number is significantly higher, the observed degree of polarization is very close to the intrinsic one, i.e. only little depolarization is possible for instance due to Faraday rotation depolarization of a foreground magnetic field.

\subsection{Faraday depolarization in the Toothbrush relic}
\label{sec:dp}

	\begin{figure}
		\centering
		\includegraphics[width=8.5cm, trim=1cm 8cm 1cm 7cm, clip=true]{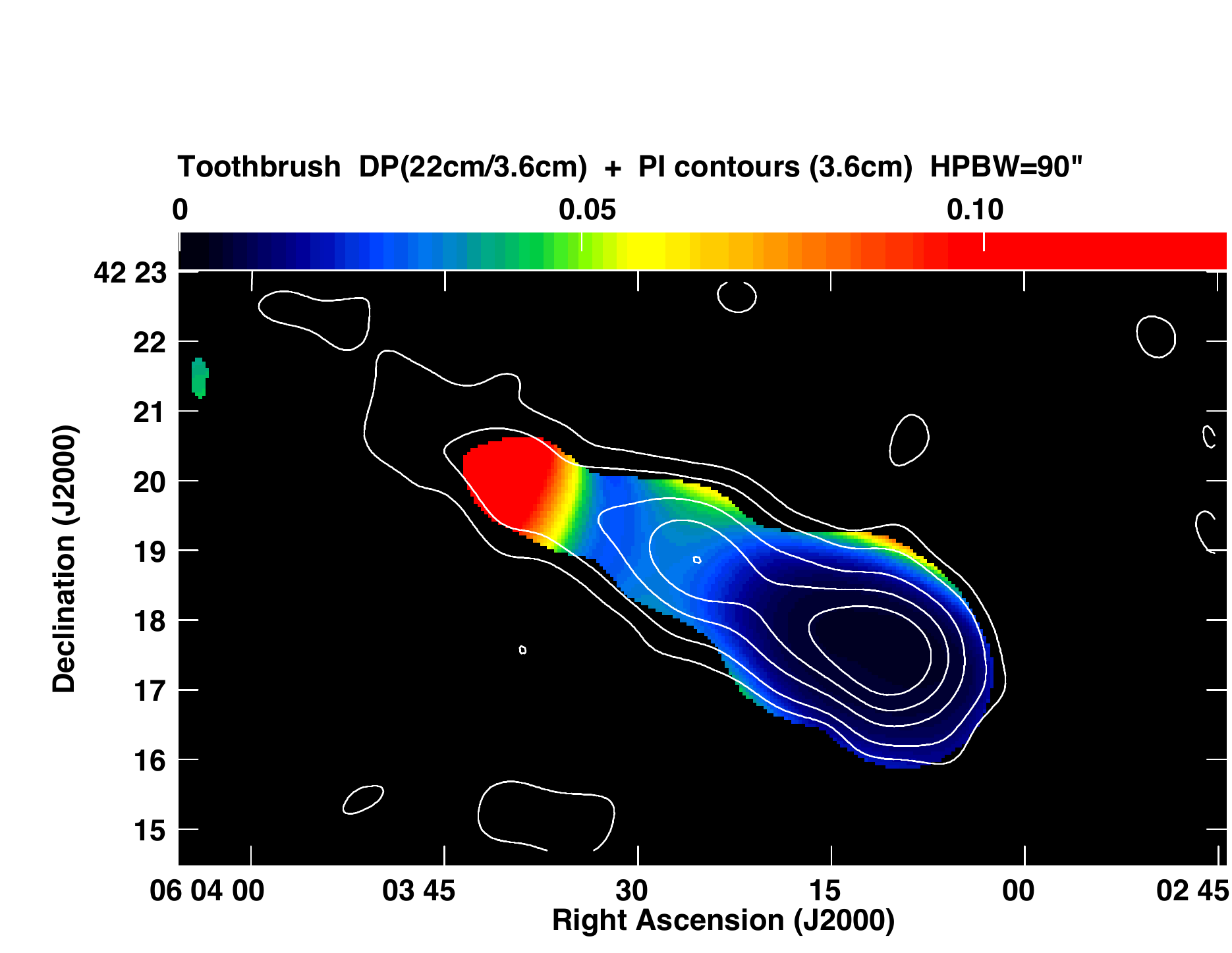}
		\caption{Depolarization between 1.38\,GHz and 8.35\,GHz (color), defined as $DP=p_{1.38}/p_{8.35}$, of the
			Toothbrush relic with a beam size $90\arcsec\times90\arcsec$, overlaid with contours of polarized intensity
			at 8.35\,GHz. Contours are drawn at levels of [1, 2, 4, 6, 8]\,$\times\,0.39$\,mJy\,/\,beam.}
		\label{fig:DP}
	\end{figure}

	Wavelength-dependent Faraday depolarization at GHz frequencies needs strong magnetic fields and/or high densities of thermal electrons \citep{1998MNRAS.299..189S,2011MNRAS.418.2336A}. Strong gradients in Faraday rotation can also cause depolarization. Clear evidence for the existence of magnetic fields in the ICM comes from radio emission in radio relics and halos, while the morphology of the fields is still unknown. The polarization of radio relics could, for instance, be caused either by large-scale ordered fields or by small-scale anisotropic fields, as discussed above. The alignment of the polarization orientation (B--vectors) with the shock surface favors small-scale anisotropic (compressed) fields. Because of the small path length along the line of sight within the relic, these fields are less efficient to depolarize than large-scale fields in the ICM or in the intergalactic medium. Hence we assume in the following that Faraday depolarization (if any) occurs in the medium in front of the relic.

	The dispersion in $\text{RM}$ within the telescope beam, $\sigma_\text{RM}$, is a useful parameter to characterize Faraday depolarization \citep{1966MNRAS.133...67B,1998MNRAS.299..189S,2011MNRAS.418.2336A}. In the case of a turbulent medium between the observer and the emitting volume the wavelength dependent depolarization depends on the Faraday dispersion according to
	\begin{equation}
		 p_\lambda
		 =
		 p_0 \, \exp ( -2\,\sigma_\text{RM}^2\,\lambda^4 ) \, ,
	\end{equation}
	where $p_0$ is the intrinsic (wavelength-independent) degree of polarization.

	Interestingly, for the Sausage relic, and for the relics in ZwCl\,0008 and in Abell\,1612, we find no evidence for depolarization. Only for the Toothbrush relic the analysis of our images indicates wavelength-dependent depolarization (Table~\ref{tab:fluxes}), with an average depolarization factor of $\text{DP}_{4.85\,\mathrm{GHz}}^{8.35\,\mathrm{GHz}}=p_{4.85\,\mathrm{GHz}}/p_{8.35\,\mathrm{GHz}}=0.68\pm0.10$. Strong Faraday depolarization has been found between 4.9\,GHz and 1.38\,GHz in components B1 and B2 of the Toothbrush relic by \citet{toothbrush}, which is clearly demonstrated in Figure~\ref{fig:tooth_WSRT}.

	A combination of our new Effelsberg observations at 8.35\,GHz and the data at 1.38\,GHz (WSRT) from \citet{toothbrush}, smoothed to the resolution of $90\arcsec\times90\arcsec$ of the Effelsberg image, yields a map of wavelength-dependent depolarization (Figure~\ref{fig:DP}). We find $\text{DP}_{1.38}^{8.35} < 0.03$ for component B1, $\text{DP}_{1.38}^{8.35} = 0.06\pm0.02$ for component B2, and $\text{DP}_{1.38}^{8.35} = 0.56\pm0.11$ for component B3, which requires $\sigma_\text{RM} > 28$\,rad\,m$^{-2}$ for B1, $\sigma_\text{RM}=25\pm2$\,rad\,m$^{-2}$ for B2, and $\sigma_\text{RM}=11\pm2$\,rad\,m$^{-2}$ for B3.

	The maps of the Toothbrush relic at 4.85\,GHz and 8.35\,GHz at a common resolution of $159\arcsec$ show depolarization of $\text{DP}_{4.85}^{8.35}=0.63\pm0.15$ for component B1, which allows us to constrain its Faraday dispersion to $\sigma_\text{RM}=134\pm35$\,rad\,m$^{-2}$. No significant depolarization between 4.85\,GHz and 8.35\,GHz is detected for components B2 and B3, as expected for small values of $\sigma_\text{RM}$.

	In summary, Faraday dispersion along the Toothbrush relic reveals a clear trend, with the largest value of $\sigma_\mathrm{RM} \simeq 130$\,rad\,m$^{-2}$ for the component B1 which is located at region with dense gas, a moderate one ($\sigma_\mathrm{RM} \simeq 25$\,rad\,m$^{-2}$) for B2, and a low one ($\sigma_\mathrm{RM} \simeq 10$\,rad\,m$^{-2}$) for the component B3 located at the outskirts of the cluster.

	In a simplified model \citep{1998MNRAS.299..189S} Faraday dispersion is described as
	\begin{equation}
		\sigma_\mathrm{RM}
		=
		\sqrt{1/3} \,\, 0.81 \, \langle n_\text{e} \rangle \, B_\mathrm{turb}
		 \,(L \, l / f))^{0.5},
	\end{equation}
	where $ \langle n_\mathrm{e} \rangle$ is the average thermal electron density of the ionized gas along the line of sight (in $\ccm$), $B_\mathrm{turb}$ the strength of the turbulent (or tangled) field (in $\muG$), $L$ the path length through the thermal gas (in pc), $l$ the turbulence scale (in pc), and $f$ the volume filling factor of the Faraday-rotating gas. Typical values for the intracluster medium of $B_\text{turb}=1.5\muG$ \citep{2004IJMPD..13.1549G}, $ \langle n_\mathrm{e} \rangle \,=10^{-3}\ccm$, $L=2.5$\,Mpc (the cluster extent), $l=10$\,kpc \citep[e.g.][]{2004A&A...424..429M}, and $f=0.5$ yield $\sigma_\mathrm{RM}\approx130\radm$ which depolarize component B1. A decrease in the product $(B_\mathrm{turb}  \langle n_\mathrm{e} \rangle (L\,l)^{0.5})$ by a factor of about 6 can explain the depolarization of component B2 ($\sigma_\mathrm{RM} \simeq 25$\,rad\,m$^{-2}$). This component has a distance to B1 of about 0.7\,Mpc and is located at the edge of the X-ray emitting cluster. A further decrease by a factor of about 2.5 is needed to explain the weak depolarization for component B3 ($\sigma_\mathrm{RM} \simeq 10$\,rad\,m$^{-2}$) that is located at a distance of about 1.2\,Mpc to B1, the outskirts of the region of hot gas. A typical value for the thermal electron density at a radius of a cumulative overdensity of 500, $R_{500}$, is $ n_\mathrm{e} \approx 10^{-4} \, \ccm$, see e.g. \citet{2008A&A...487..431C}. If the values adopted for the strength of the turbulent field and the turbulent scale reflects the properties of the ICM, this suggests that the component B1 of the Toothbrush is depolarized by the ICM significantly within $R_{500}$, B2 is depolarized by the ICM roughly located at  $R_{500}$, and B3 is depolarized only by medium residing in the outskirts of the cluster. Hence, the ``brush'' is located to some extent behind the cluster.


\subsection{Faraday Rotation Measures}
\label{sec:rm}

	\begin{figure*}
	\centering
	\includegraphics[width=8cm, trim=1cm 6cm 1cm 3cm, clip=true]{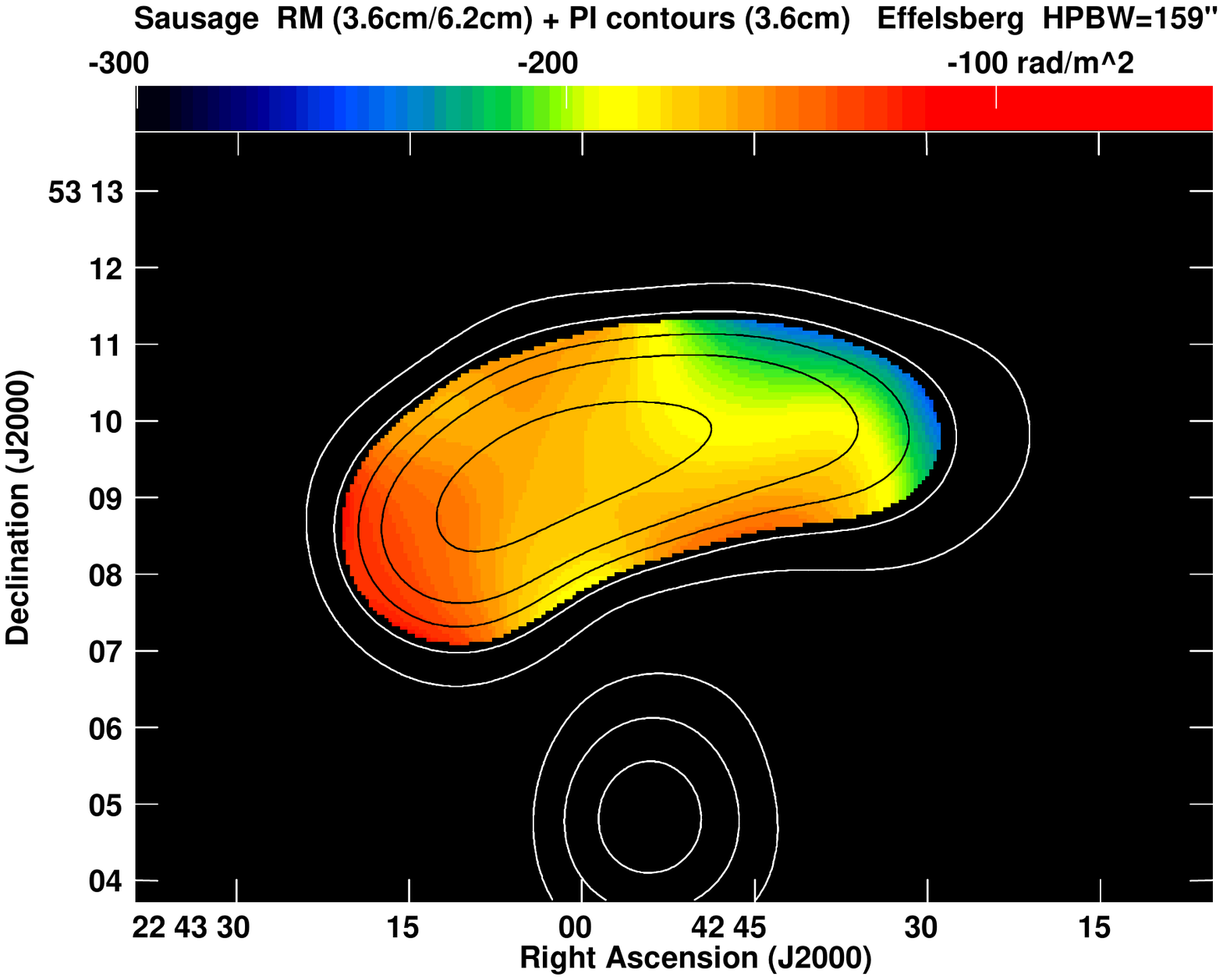}
	   \includegraphics[width=8.5cm, trim=1cm 6cm 0.5cm 3cm, clip=true]{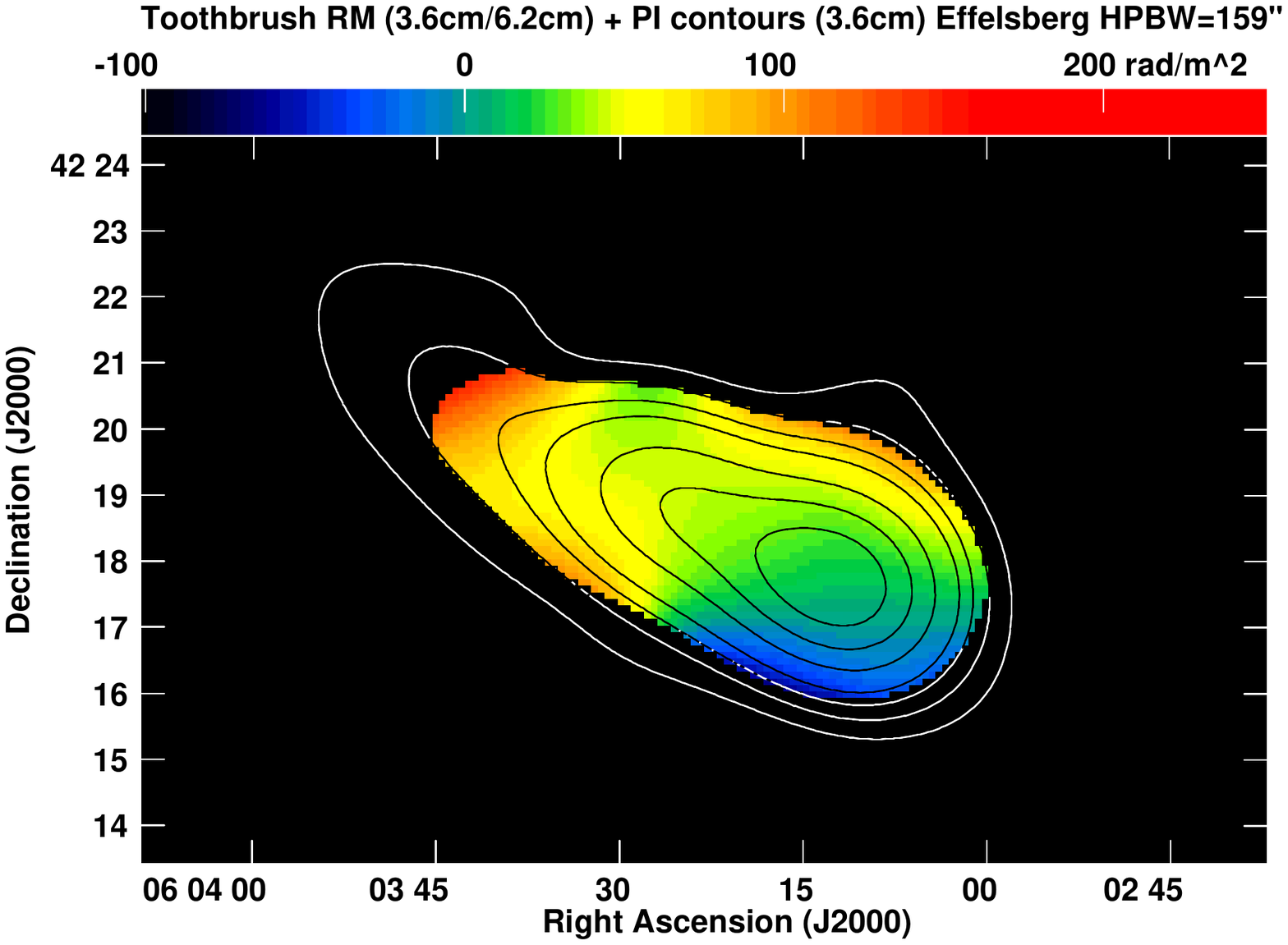}
	     \caption{Maps of Faraday rotation measure (color, in rad\,m$^{-2}$) of the Sausage relic (\textit{left}) and the Toothbrush relic (\textit{right}),
	     computed for regions where the polarized intensity exceeds ten times the rms noise $\sigma_{QU}$ in U and Q at
	     both frequencies. The contours show the polarized intensity at 8.35\,GHz at [1, 2, 3, 4, 6]
	     \,$\times\,0.3$\,mJy\,/\,beam for the Sausage relic (\textit{left}) and at [1, 2, 3, 4, 6, 8, 10]
	     \,$\times\,0.5$\,mJy\,/\,beam for the Toothbrush relic (\textit{right}). The beam sizes are $159\arcsec\times159\arcsec$.
The $\text{RM}$ of $+532 \pm 39\radm$ of source D south of the Sausage relic is beyond the color range and
is not shown in the left figure.
}
	     \label{fig:RM}
	\end{figure*}

	\begin{figure*}
	\centering
	\includegraphics[width=8.5cm, trim=1cm 1cm 1cm 2cm, clip=true]{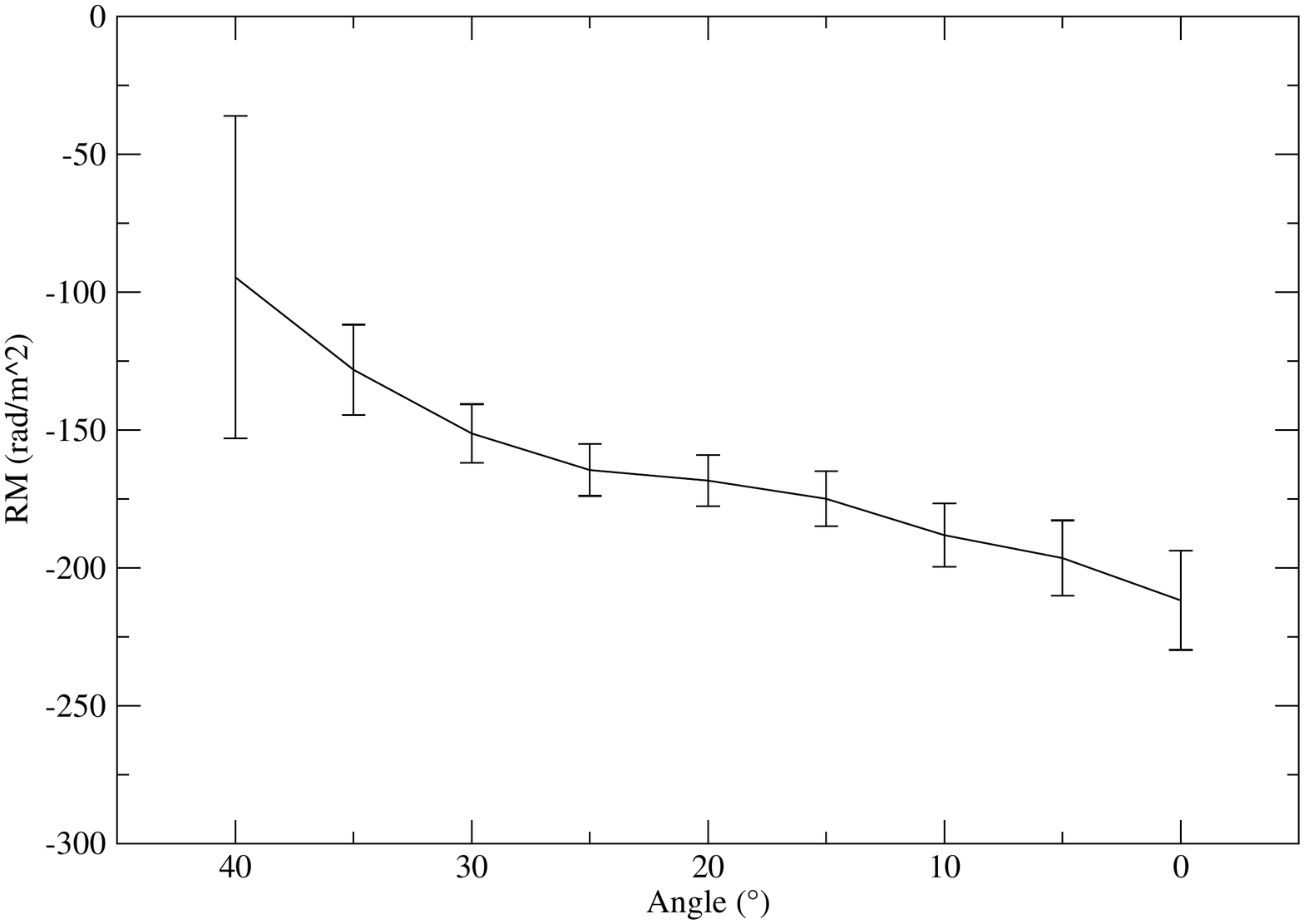}
	   \includegraphics[width=8.5cm, trim=1cm 1cm 1cm 2cm, clip=true]{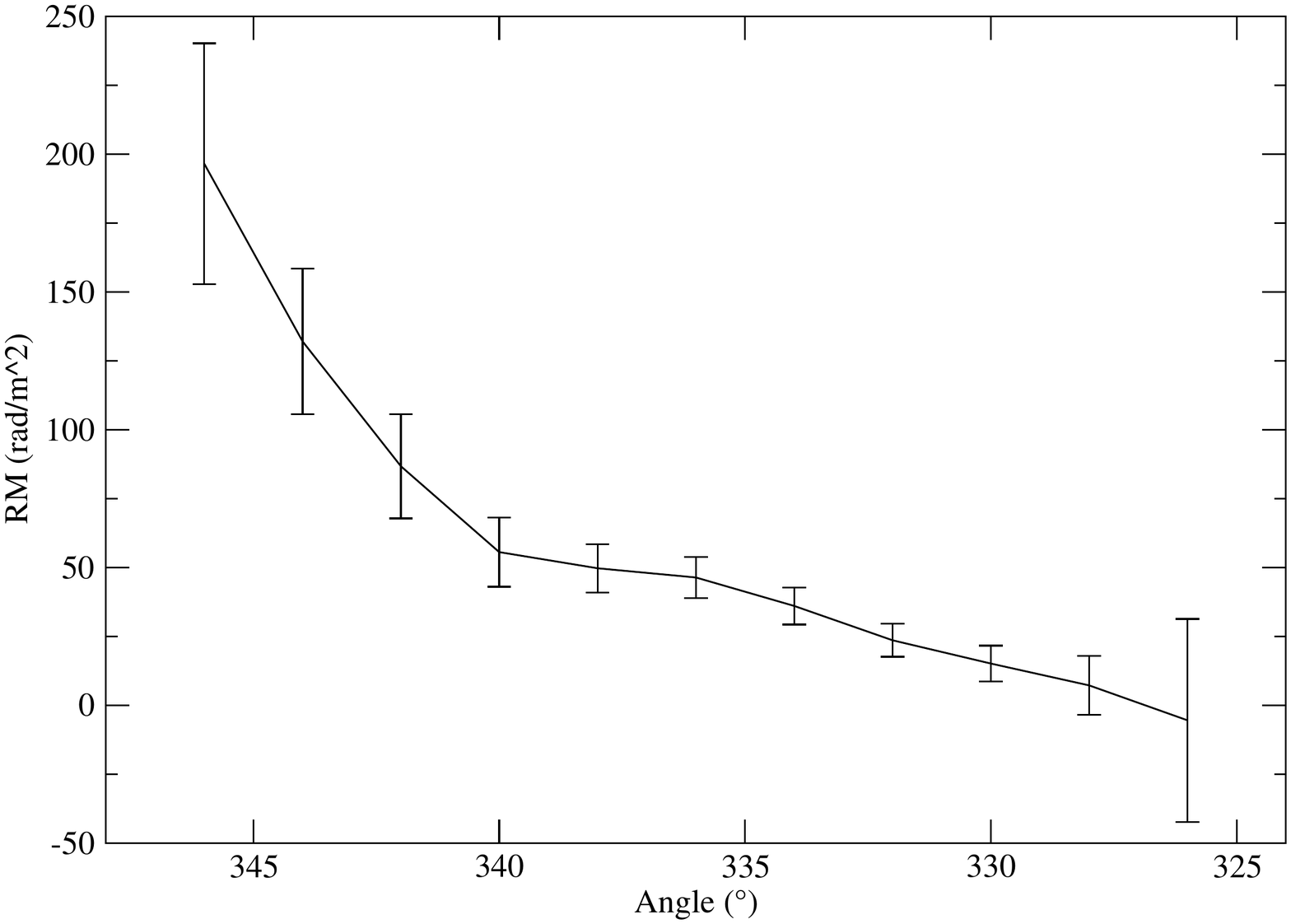}
	     \caption{Profile of not foreground subtracted rotation measures of the Sausage relic (\textit{left}) and the Toothbrush relic
	     (\textit{right})   in sectors of $1\arcmin$ width (corresponding to $5\degr$ and $2\degr$
	     azimuthal angle along the ring for the Sausage and Toothbrush relics, respectively) and of $3\arcmin$
	     thickness in radius. The azimuthal angle on the horizontal axis is measured counter-clockwise from the
	     north-south direction. The ring for each relic is centered at the approximate cluster center and follows
	     the shape of the relic. The error bars are computed from the rms noise of $U$ and $Q$ in each sector.}
	     \label{fig:RM_sector}
	\end{figure*}
	
	\begin{table}
		\caption{Mean $\text{RMs}$ between 4.85\,GHz and 8.35\,GHz and in the Galactic foreground}
		\begin{tabular}[t]{lcc}
			\toprule\toprule
			Name (component)			&RM &RM$_\mathrm{fg}$	\\
			& ($\radm$) & ($\radm$) \\
			(1)      					&(2)		&(3) \\
			\midrule
			Sausage (east)				& $-147\pm10$ & $-73\pm70$\\
			Sausage (middle)			& $-172\pm10$ \\
			Sausage (west)				& $-181\pm15$ \\
			CIZA J2242$+$53 (Source D)	& $+532\pm39$ \\
			Toothbrush (east, comp B3)	& $+48\pm8$   & $+16\pm28$\\
			Toothbrush (west, comp B1)	& $+9\pm6$    \\
			ZwCl\,008						& $-63\pm32$  & $-2\pm36$\\
			A\,1612					    & $+8\pm22$   & $-5\pm6$\\
			\bottomrule
		\end{tabular}
		\tablefoot{(1) Source name and component; (2) Rotation Measures (RMs) between 4.85\,GHz and 8.35\,GHz without foreground correction;
			(3) foreground $\text{RM}_\mathrm{fg}$ estimated from the catalog of \citet{2012A&A...542A..93O}.}
		\label{tab:RM}
	\end{table}

	As we observe only at two frequencies, the rotation measure was estimated assuming a linear dependency of the polarization angle $\chi$ and the wavelength $\lambda$:
	\begin{equation}
		\text{RM}
		=
		\frac{\chi_1-\chi_2}{\lambda^2_1-\lambda^2_1}
	\end{equation}
	This method gives average Faraday rotation angles, not corrected for redshift. Information on the complexity of the relics in Faraday space would need broadband polarimetric data and application of RM Synthesis (Section~\ref{sec:conclusion}).

	To obtain the rotation measure maps we convolved the Stokes U and Q maps at 8.35\,GHz to the resolution of the 4.85\,GHz maps of $159\arcsec$. The final $\text{RM}$ maps were made using the low-resolution maps of polarization angles, clipped below a signal-to-noise ratio of $S_\text{p}=\text{PI}/\sigma_\text{UQ}=10$, where $\sigma_\text{UQ}$ is the rms noise in U and Q at both frequencies (see Table \ref{tab:observation3cm} and \ref{tab:observation6cm}). The error in $\text{RM}$ is $\Delta_\text{RM}=(\sqrt{2}\,S_\text{p}\,\Delta\lambda^2)^{-1}$, which is $\Delta_\text{RM}\simeq 28\radm$ for $S_\text{p}=10$. The $\pm n\,\pi$ ambiguity of polarization angles corresponds to a large $\text{RM}$ ambiguity of $\pm n \times 1241\radm$ that can be excluded for our data.

	The results of the $\text{RM}$ analysis are shown in Figure~\ref{fig:RM} for the two bright radio relics. No $\text{RM}$ maps are shown for the two faint relics because $\text{RM}$ can be measured only in a small region for these candidates.

	The mean $\text{RMs}$ between 4.85\,GHz and 8.35\,GHz in several regions are summarized in Table~\ref{tab:RM}, computed from the average Stokes U and Q values at both frequencies. The uncertainties in the mean $\text{RM}$ were estimated from error propagation, using the rms noise in U and Q at both frequencies.

	For the Sausage radio relic in CIZA\,J2242 we find $\text{RMs}$ between $-150$\,rad\,m$^{-2}$ and $-180$\,rad\,m$^{-2}$. The values are in agreement with the average value of $-140$\,rad\,m$^{-2}$ obtained at 1.2--1.8\,GHz \citep{2010Sci...330..347V}. If regular fields within the relic would generate the large observed values of RM, total depolarization by differential Faraday rotation has to occur at 1.2--1.8\,GHz \citep[see e.g.][]{2011MNRAS.418.2336A},
which is not the case. As the anisotropic magnetic fields in relics do not give rise to Faraday rotation, the $\text{RM}$ probably originates in the foreground. Located near to the plane of the Milky Way at Galactic coordinates of $l=104\degr$, $b=-5\degr$, strong Faraday rotation is expected. The interactive catalogue based on \citet{2012A&A...542A..93O} gives a mean value of $\text{RM} = -73 \pm 70\radm$ around the relic position (Table~\ref{tab:RM}), which is consistent with our result. A $\text{RM}$ gradient of about $12\radm$ per arcminute along the relic (Figure~\ref{fig:RM_sector} left) is indicated, which is much larger than the predictions of the $\text{RM}$ foreground by \citet{2009A&A...507.1087S}. A refined run of this simulation towards the Sausage relic at a resolution of $2.5\arcmin$ gives a $\text{RM}$ gradient of only about $10\radm$ per degree (Reich, priv. comm.).

	Source D in CIZA\,J2242$+$53 is a bright head-tail radio galaxy \citep{CIZA} for which we find a very high $\text{RM}$ of $+532 \pm 39\radm$ (Figure~\ref{fig:RM} left, bottom) that most probably originates from the radio lobes of the galaxy itself.

	The average $\text{RM}$ of the Toothbrush relic is also consistent with the value of the Galactic foreground (Table~\ref{tab:RM}). A $\text{RM}$ gradient of about $15\radm$ per arcminute along the relic is indicated. Again, such a gradient is much larger than the Galactic foreground can produce. \citet{toothbrush} applied RM Synthesis to the data in the 1.38\,GHz band and found an even larger $\text{RM}$ difference between components B1 ($-53\radm$) and B2 ($+55\radm$). Strong Faraday rotation is also obvious when comparing the apparent B--vectors between 1.38\,GHz and 4.9\,GHz (Figure~\ref{fig:tooth_WSRT}). No $\text{RM}$ map was computed between these frequencies because the small $\text{RM}$ ambiguity of $\pm n \times 72\radm$ does not allow us to obtain reliable $\text{RM}$ values.
	

	For ZwCl and A\,1612 there were no $\text{RM}$ values published so far. While the small $\text{RMs}$ for A\,1612 are consistent with its high Galactic latitude of $b=+60\degr$, RM$\simeq -60\radm$ for ZwCl deviates from the average value estimated from the catalog by \citet{2012A&A...542A..93O}.
	
	For the large radio relic in Abell\,2256, a high resolution $\text{RM}$ map obtained with JVLA observations in the frequency range from 1 to 8\,GHz has been published \citep{2014ApJ...794...24O}. The map shows large coherent $\text{RM}$ patches with $\text{RM}$ values ranging from about -70 to about -20\,$\radm$. The $\text{RM}$ gradients are significantly larger than expected for the Galactic foreground.

	The generation of $\text{RM}$ and $\text{RM}$ gradients needs large-scale regular magnetic fields that preserve their directions over the extent of the relics of several Mpc, much larger than the coherence length of the turbulent (or tangled) field. A large-scale $\alpha-\Omega$ dynamo cannot operate as in spiral galaxies because galaxy clusters do not have organized rotation. The magneto-thermal instability may generate radially ordered fields in the ICM \citep{2010NatPh...6..520P}, but without coherent directions.

	Simulations indicate that the strengths of large-scale intergalactic magnetic fields are about 100\,nG around galaxy clusters and about 10\,nG in filaments of the cosmic web \citep[e.g.][]{2008Sci...320..909R}. With electron densities of typically $ \langle n_\mathrm{e} \rangle \,= 10^{-4}\ccm$ at $R_{500}$ \citep{2008A&A...487..431C}, a pathlength through the regular field (coherence length) of 1--10\,Mpc is needed to achieve $10\radm$. Large-scale filaments with coherent RMs were found in the relic of Abell~2256 \citep{2014ApJ...794...24O}. The observed lengths of several 100\,kpc are limited by the extent of synchrotron emission, while the magnetic filaments could be much larger. Recent MHD simulations indicate magnetic filaments on such scales (Vazza, priv. comm.). Our observations give evidence that such large-scale intergalactic fields may exist around galaxy clusters.

	As a note of caution, we have to admit that the RMs in Figure~\ref{fig:RM} are based on measurements of polarization angles in only two relatively narrow frequency bands. If the polarization angle is not proportional to $\lambda^2$, e.g. due to Faraday depolarization or internal structure of the emitting and Faraday-rotating source (a complex distribution of Faraday rotating and synchrotron emitting media), two-point RMs are not reliable. The first issue can be excluded because Faraday depolarization is small for all relics observed in this work (except for component B1 of the Toothbrush relic). Internal structure within the relic cannot be excluded and indeed is a possible scenario \citep{2014ApJ...785....1B}. Polarization observations over a wide range in $\lambda^2$ and application of RM Synthesis \citep{2005A&A...441.1217B} are needed to measure the Faraday spectrum that contains information about the emitting region. Furthermore, signatures of the turbulent or tangled field should become visible with measurement with high resolution in Faraday space as a ``Faraday forest'' of many components in the Faraday spectrum \citep{2012A&A...543A.113B}.

\section{Conclusions and discussion}
\label{sec:conclusion}

	Single-dish radio telescopes like the Effelsberg 100-m are ideal instruments to confirm relic candidates or search for new cluster relic candidates through detecting their polarization, as long as the map area is sufficient to achieve reliable baselines, to determine meaningful rms noise values and hence to measure the total and polarized flux densities with small uncertainties. Although the resolution is limited due to the large beam size, magnetic fields in cluster relics are highly ordered, so that the degree of polarization only weakly decreases with increasing beam size. Even relics with angular extents smaller than the telescope beam still reveal significant polarization, if the curvature of the shock front is small within the beam.
	
	Galaxy cluster relics are among the most highly polarized sources on sky. In the two prominent Mpc-sized radio relics, the ``Sausage'' (in the cluster CIZA\,J2242$+$53) and the ``Toothbrush'' (1RXS\,06$+$42), we found maximum polarization degrees of about 50\,\% at 8.35\,GHz, which are among the highest fractions of linear polarization detected in any extragalactic radio source, indicating strong shocks (high Mach numbers). We detected polarization also from the eastern relic in the cluster ZwCl\,0008$+$52 and from a relic in Abell\,1612. The maximum degrees of polarization of the latter two relics are smaller (20--30\,\%) because either their shock strengths are lower and/or the relics are inclined to the sky plane.

	The radio spectra below 8.35\,GHz can be well fitted by single power laws for all four relics. Possible spectral breaks toward higher frequencies \citep{2016MNRAS.455.2402S} need confirmation by further observations and corrections for the SZ  effect \citep{2016A&A...591A.142B}. The flat average spectral indices of 0.9 and 1.0 for the Sausage and the Toothbrush relics indicate that models describing the origin of relics have to include effects beyond the assumptions of diffuse shock acceleration. The steep radio spectra for the relics in ZwCl\,0008$+$52 and Abell\,1612 give additional evidence for low Mach numbers of $\approx 2.4$.
	

	The total field strengths of the relics are in the range of 2.4--4.2\,$\mu$G, assuming equipartition between the energy densities of cosmic rays and a proton-to-electron ratio of 100. These field strengths are in accord with upper limits of the inverse Compton (IC) emission. It has been proposed that at merger shocks electrons are accelerated via shock drift acceleration \citep{2014ApJ...794..153G}. This process is only possible for a high plasma--$\beta$ upstream of the shock ($\beta \gtrsim 20$). If only electrons are accelerated the proton-to-electron ratio is 0 and the corresponding estimates of magnetic field strengths become 0.7--1.2\,$\mu$G, fulfilling the high plasma--$\beta$ requirement. However, these fields are in tension with the limits derived from IC emission. Deeper IC measurements of the cosmic-ray electron density are required to solve this discrepancy.
			
	We observed Abell\,1612 with the Chandra telescope. The complex X-ray surface brightness distribution reveals an ongoing major merger after first core passage and supports the classification of the diffuse emission as radio relic.

	No wavelength-dependent Faraday depolarization is detected between 4.85\,GHz and 8.35\,GHz, except for one component of the Toothbrush relic, confirming the low density of thermal electrons within the relic and within the ICM in front of the relic. Faraday depolarization is significant for the Toothbrush relic between 1.38\,GHz and 8.35\,GHz and varies with distance from the cluster center. This can be explained by a decrease in electron density and strength of the turbulent field. We conclude that measurements of Faraday depolarization at radio frequencies of several GHz, preferably with single-dish telescopes which trace the full extended emission, allow us to put constraints on the properties of the intracluster medium.

	Faraday rotation measures along the relics reveal large-scale gradients that are much larger than those expected in the Milky Way foreground. This gives evidence for large-scale intergalactic fields around the clusters.


	Polarization observations with synthesis telescopes at low or moderate frequencies covering a wide range in $\lambda^2$ and application of RM Synthesis \citep{2005A&A...441.1217B} will allow us to measure the Faraday spectrum and extract information on the internal structure of the emitting region and the turbulent medium in the cluster foreground.


\begin{acknowledgements}
The authors acknowledge support from the DFG Research Unit FOR1254.
We thank the referee for helpful comments.
We thank Alice di Vicenzo for performing a part of the observations and the operators at the Effelsberg telescope for their support.
We thank Dr.\,Aritra Basu (MPIfR) for supporting us with the interpretation of our data and for valuable suggestions to improve the paper.
We thank Wolfgang Reich (MPIfR) and Franco Vazza (Hamburger Sternwarte) for important contributions to the discussion.
We thank Felipe Andrade-Santos for providing various Chandra data reduction scripts. R.J.W. is supported by a Clay Fellowship awarded by the Harvard-Smithsonian Center for Astrophysics.
W.~R.~Forman and C.~Jones acknowledge support from the Smithsonian Institution and the Chandra HRC GTO Program.
\end{acknowledgements}

\bibliography{relics}{}
\bibliographystyle{aa}


\end{document}